\documentclass[useAMS]{mn2e}
\usepackage{graphicx}
\usepackage{epsfig}
\usepackage{amssymb}
\usepackage{lscape}
\usepackage{ulem}
\usepackage{txfonts}

\def\doh{$\Delta$O/H}

\def\logms{$\log$ (M$_{\star}$/M$_{\odot}$)}
\def\aq{ALMaQUEST}
\def\sigh2{$\Sigma_{\rm H_2}$}
\def\sigsfr{$\Sigma_{\rm SFR}$}

\def\sigstar{$\Sigma_{\star}$}
\def\dsfr{$\Delta$SFR}
\def\dsigsfr{$\Delta \Sigma_{\rm SFR}$}
\def\dsfe{$\Delta$SFE}
\def\dfgas{$\Delta f_{\rm H_2}$}
\def\fh2{$f_{\rm H_2}$}

\def\c2s{C\,{\sc ii}$^{\star}$}

\def\dsfr{$\Delta$SFR}

\title[ALMaQUEST II - Central starbursts] {The ALMaQUEST Survey: II. What drives central starbursts at z$\sim$0?}

\author[Ellison et al.] {Sara L. Ellison$^1$,  Mallory D. Thorp$^1$,  Hsi-An Pan$^2$, Lihwai Lin$^2$, Jillian M. Scudder$^3$, \newauthor Asa F. L. Bluck$^4$,  Sebastian F. S\'{a}nchez$^5$, Mark Sargent$^6$\\ 
$^1$ Department of Physics \& Astronomy, University of Victoria, Finnerty Road, Victoria, British Columbia, 
  V8P 1A1, Canada\\
  $^2$ Institute of Astronomy \& Astrophysics, Academia Sinica, Taipei 10617, Taiwan\\
  $^3$ Department of Physics and Astronomy, Oberlin College, Oberlin, Ohio, OH 44074, USA\\
  $^4$ Kavli Institute for Cosmology \& Cavendish Astrophysics, University of Cambridge, Madingley Road, Cambridge, CB3 0HA, UK\\
  $^5$ Instituto de Astronom\'{i}a, Universidad Nacional Autonoma de Mexico, A. P. 70-264, C.P. 04510, Mexico, 
  D.F., Mexico\\
  $^6$ Astronomy Centre, Department of Physics and Astronomy, University of Sussex, Brighton BN1 9QH, UK
}

\begin{document}

\maketitle

\begin{abstract}
  Starburst galaxies have elevated star formation rates (SFRs) for their stellar mass.  In Ellison et al. (2018) we used integral field unit (IFU) maps of star formation rate surface density (\sigsfr) and stellar mass surface density (\sigstar) to show that starburst galaxies in the local universe are driven by SFRs that are preferentially boosted in their central regions.  Here, we present molecular gas maps obtained with the Atacama Large Millimeter Array (ALMA) observatory for 12 central starburst galaxies at  $z\sim0$ drawn from the Mapping Nearby Galaxies at Apache Point Observatory (MaNGA) survey.  The ALMA and MaNGA data are well matched in spatial resolution, such that the ALMA maps of molecular gas surface density (\sigh2) can be directly compared with MaNGA maps at kpc-scale resolution.  The combination of \sigh2, \sigstar\ and \sigsfr\ at the same resolution allow us to investigate whether central starbursts are driven primarily by enhancements in star formation efficiency (SFE) or by increased gas fractions.  By computing offsets from the resolved Kennicutt-Schmidt relation (\sigh2\ vs. \sigsfr) and the molecular gas main sequence (\sigstar\ vs. \sigh2), we conclude that the primary driver of the central starburst is an elevated SFE.  We also show that the enhancement in \sigsfr\ is accompanied by a dilution in O/H, consistent with a triggering that is induced by metal poor gas inflow.  These observational signatures are found in both undisturbed (9/12 galaxies in our sample) and recently merged galaxies, indicating that both interactions and secular mechanisms contribute to central starbursts.
  
\end{abstract}


\section{Introduction}

The regulation of star formation in galaxies is one of the most fundamental processes in astronomy.  Insight into the mechanisms that drive star formation on both local and global scales has seen tremendous progress in the last decade, thanks to a combination of large surveys that measure the global properties of millions of galaxies, complemented with kpc-scale mapping in gas and stars on smaller samples.  These approaches are highly complementary, combining vast statistics of global metrics, with the detailed dissection of the most nearby galaxies.

From massive contemporary galaxy surveys, such as the Sloan Digital Sky Survey (SDSS), we have learned that correlations exist between properties such as total stellar mass, global star formation rate (Brinchmann et al. 2004; Noeske et al. 2007; Salim et al. 2007), metallicity (Tremonti et al. 2004; Kewley \& Ellison 2008), local environment (Peng et al. 2010; Woo et al. 2013; Lin et al. 2014), internal structure (Wuyts et al. 2011; Bluck et al. 2014) and gas content (Catinella et al. 2010; Saintonge et al. 2011).  However, there is considerable scatter within these global scaling relationships, indicating a more complex interplay of physical processes in the long-term assembly of galaxies.  For example, scaling relations between two variables may have a higher order dependence on a third parameter.  A well-known example of this is the reduction in scatter in the mass-metallicity relation when considering either star formation rate (e.g. Ellison et al. 2008a; Lara-Lopez et al. 2010; Mannucci et al. 2010; Salim et al. 2014) or gas fraction (Bothwell et al. 2013, 2016; Brown et al. 2018).  Understanding the underlying drivers of star formation regulation therefore often requires multi-parameter analyses (e.g. Peng et al. 2010; Woo et al. 2015; Bluck et al. 2016; Teimoorinia, Bluck \& Ellison 2016; Bluck et al. 2019).  The goal of the work presented here is to identify why some galaxies are driven above the global main sequence, with a particular focus on those that host central starbursts.

Ultimately, the creation of stars from molecular gas clouds is a process that operates on sub-kpc scales, well below the spatial resolution of surveys such as the SDSS, or their gas content counterparts (e.g. Saintonge et al. 2017; Catinella et al. 2018).  In order to reach beyond the clues that come from global scaling relationships, it is therefore necessary to map the gas, stellar and metal content on kpc scales (or below).  Initial progress on this front was made using single dish CO observations for a few tens of the closest galaxies, such as the Heterodyne Receiver Array CO Line Extragalactic Survey (HERACLES; Leroy et al. 2009).  However, local samples are limited in the diversity of galaxies that they can investigate.  Creating larger samples of diverse galaxy populations for which the stars and gas can be mapped on kpc-scales has been enabled by two complementary advances in observational facilities.  First, large integral field unit (IFU) galaxy surveys now routinely map stars and ionized gas for hundreds or thousands of nearby galaxies at kpc resolution (e.g. S\'{a}nchez et al. 2012; Bundy et al. 2015).  Second, CO follow-up using interferometers can match the spatial scale of local IFU surveys, e.g. the Extragalactic Database for Galaxy Evolution Calar Alto Legacy Integral Field Area (EDGE-CALIFA) survey (Bolatto et al. 2017), providing resolution matched maps of molecular gas for larger samples that are more representative of the low redshift universe.

The combination of stellar mass, star formation rate (SFR) and gas maps has revealed scaling relationships at the kpc-scale that are starting to shed light on the origins of the global galaxy correlations and provide a physically motivated view of star formation regulation.  From the IFU surveys alone (which leverage the full potential of sample size, containing millions of spaxels), it has been found that the stellar mass surface density (\sigstar) correlates tightly with the star formation rate surface density (\sigsfr) (the so-called `resolved star forming main sequence', rSFMS: S\'{a}nchez et al. 2013; Cano-D\'{i}az et al. 2016; Hsieh et al. 2017), as well as with metallicity (Barrera-Ballesteros et al. 2016; Ellison et al. 2018) on kpc scales.  Likewise, from CO mapping surveys, the long-known relationship between molecular gas surface density (\sigh2) and \sigsfr, also known as the Kennicutt-Schmidt relation (Schmidt 1959; Kennicutt 1989, 1998b) has been established with exquisite sensitivity and resolution (e.g. Bigiel et al. 2008; Leroy et al. 2008; Schruba et al. 2011), as well as the molecular gas main sequence relation between \sigstar\ and \sigh2\ (Wong et al. 2013; Lin et al. 2019).  The availabilty of dense gas (as traced by HCN) maps for small samples of galaxies reveals additional kpc-scale dependences, such a positive correlation between the dense gas fraction and \sigstar\ (Usero et al. 2015; Bigiel et al. 2016; Gallagher et al. 2018; Jimenez-Donaire et al. 2019).

Beyond these basic scaling relationships, the combination of larger samples and multi-scale diagnostics has revealed a more complex interplay between gas and star formation.  For example, although the Kennicutt-Schmidt relation is well represented by a linear slope in nearby star-forming disks (e.g. Bigiel et al. 2008, 2011; Leroy et al. 2008, 2013; Schruba et al. 2011; Rahman et al. 2012), indicating that star formation efficiencies (SFE = \sigsfr/\sigh2) are largely universal, bars and mergers can affect the central depletion time (e.g. Bolatto et al. 2017; Utomo et al. 2017; Chown et al. 2019).  Likewise, although SFR correlates well with dense molecular gas mass on scales from molecular clouds to galaxies (e.g. Gao \& Solomon 2004; Wu et al. 2005; Lada et al. 2010), the significant scatter in this relation, and anti-correlation of the star formation efficiency of dense gas on \sigstar, point to local environmental regulation of star formation (Usero et al. 2015; Bigiel et al. 2016; Gallagher et al. 2018; Jimenez-Donaire et al. 2019).  As a result, there is significant galaxy-to-galaxy variation of the rSFMS and the SFE can vary by an order of magnitude or more within a given galaxy (e.g. Bemis \& Wilson 2019; Tomicic et al. 2019; Vulcani et al. 2019a).

Starburst galaxies are well known to deviate from global galaxy scaling relations.  Defined as having high SFRs for their stellar mass (i.e. high global specific SFRs), starburst galaxies are found to have low metallicities (Hoopes et al. 2007; Ellison et al. 2008a), high global gas fractions (Saintonge et al. 2016),  high star formation efficiencies (Daddi et al. 2010; Genzel et al. 2010; Saintonge et al. 2011, 2012) and enhanced dense gas star formation efficiencies (Gracia-Carpio et al. 2008; Garcia-Burillo et al. 2012).   Based on spatially resolved IFU data from the Mapping Nearby Galaxies at Apache Point Observatory (MaNGA) survey, Ellison et al. (2018) demonstrated that $z \sim 0$ starburst events are preferentially occuring in the central regions of galaxies located above the global SFMS (see also Wang et al. 2019).   Local gas scaling relations indicate that these high central SFRs could be caused by high gas surface densities (e.g Bigiel et al. 2008, 2011; Leroy et al. 2008, 2013; Schruba et al. 2011), elevated central SFEs (e.g. Leroy et al. 2013; Bolatto et al. 2017; Utomo et al. 2017) or environmental effects linked to the radial profile of \sigstar\ (Usero et al. 2015; Bigiel et al. 2016; Gallagher et al. 2018; Jimenez-Donaire et al. 2019). The goal of the work presented here is to distinguish between these scenarios.

There is an abundant literature of studies aiming to isolate either gas fraction or SFE as the cause of starbursts based on \textit{global} metrics.  For example, it has long been known that molecular gas SFEs are elevated for luminous (highly star-forming) galaxies and mergers (e.g. Solomon \& Sage 1988; Solomon et al. 1997; Bryant \& Scoville 1999; Gao \& Solomon 2004), and there appears to be a general correlation between \textit{global} SFE and distance above the main sequence both locally and at high $z$ (e.g. Saintonge et al. 2012; Sargent et al. 2014; Genzel et al. 2015; Silverman et al. 2015, 2018; Tacconi et al. 2018).   However, other works have suggested that enhanced gas reservoirs are the key to regulating SFRs (Scoville et al. 2016; Lee et al. 2017).  Indeed, the gas fractions of interacting galaxies (a process known to trigger central star formation, e.g. Ellison et al. 2013; Thorp et al. 2019; Pan et al. 2019) are a factor of 2--3 in excess of isolated galaxies (Ellison et al. 2015; Ellison, Catinella \& Cortese 2018; Pan et al. 2018; Violino et al. 2018; Lisenfeld et al. 2019). Moreover, mergers are found to have high column densities of both atomic and molecular gas in their centres (Ueda et al. 2014; Bolatto et al. 2017; Espada et al. 2018; Dutta et al. 2018, 2019).  In practice, both gas availability, and the efficiency of converting that gas to stars may play a role (Saintonge et al. 2012, 2016; Combes et al. 2013; Lin et al. 2017; Scoville et al. 2017; Elbaz et al. 2018).  For example, Gao \& Solomon (2004) have pointed out that while many luminous infra-red galaxies have enhanced SFEs, there is a population of these prodigious star-formers with normal SFRs for their gas mass.  Gao \& Solomon (2004) concluded that, in this latter population, the high SFRs are due to massive gas reservoirs, but otherwise follow scaling relations and hence might not be considered 'true' starbursts.  

Further insight into the various possible drivers of starbursts may be gained by investigating the interplay of gas and stars on \textit{local} (kpc) scales, to avoid averaging properties over the entire galaxy which may mask the true physical processes at play.  Unfortunately, representative studies of kpc-scale gas and stars such as HERACLES (Leroy et al. 2009) and EDGE-CALIFA (Bolatto et al. 2017) are still too small to include sufficient numbers of rare starbursts to permit their systematic study.  Although some previous work has specifically targeted starburst samples (e.g. Gao \& Solomon 2004; Garcia-Burillo et al. 2012), these global studies do not permit kpc-scale studies of the inner regions where the starburst is occuring, and can not account for the important role of local environment (Usero et al. 2015; Bigiel et al. 2016; Gallagher et al. 2018).  In this paper, we present a kpc-scale study of 12 central starbursts selected from the MaNGA survey, with follow-up ALMA observations of CO (1-0) matched in spatial resolution in order to conduct a multi-variable kpc-scale investigation of the primary driver(s) of the central starbursts.

In Section \ref{data_sec} we describe the MaNGA and ALMA data used in this work, as well as metrics used to define deviations from local scaling relations.  Section \ref{results_sec} contains our main results, based on the azimuthally averaged profiles of starburst galaxy properties as well as a spaxel-by-spaxel correlation analysis.  Our results are discussed in Section \ref{discuss_sec} and our conclusions are summarized in Section \ref{summary_sec}.

\smallskip

We adopt a cosmology in which H$_0$=70 km/s/Mpc, $\Omega_M$=0.3, $\Omega_\Lambda$=0.7.

\section{Data}\label{data_sec}

\subsection{MaNGA}

The MaNGA data used in this paper make use of the public PIPE3D (S\'{a}nchez et al. 2016a, 2016b) data products from the SDSS Data Release 15 (DR15), with some additional quantities computed by us.  Global values used in this paper, such as total stellar mass, total SFR and redshift, are taken from the PIPE3D Value Added Catalog (VAC), provided with the DR15 release (S\'{a}nchez et al. 2018).  We also make use of the public PIPE3D stellar mass surface densities (\sigstar) and line fluxes, the latter of which we correct for internal extinction assuming a Milky Way extinction curve (Cardelli, Clayton \& Mathis 1989) and an intrinsic H$\alpha$/H$\beta$ = 2.85.  Surface densities of star formation rate (\sigsfr) are computed by using the extinction corrected H$\alpha$ luminosities in equation 2 from Kennicutt (1998a), assuming a Salpeter initial mass function (to be consistent with the PIPE3D products).    Metallicities are also computed by us using the extinction corrected emission line fluxes with the Marino et al. (2013) O3N2 calibration.  Finally, we compute inclination corrected galactocentric radii from the V-band centre of the galaxy using the b/a axial ratios from single Sersic fits provided in the NASA Sloan Atlas (NSA) catalog.  Inclination corrections are also applied to \sigsfr\ and \sigstar.  In the analysis presented here, we only use MaNGA spaxels that are classified as star forming according to the Kauffmann et al. (2003) diagnostic criteria, requiring that the S/N in each of the emission lines required for that diagnostic is $\ge$ 3, and that the H$\alpha$ equivalent width exceeds 6 \AA.  There are $\sim$ 1.5 million star-forming spaxels in all of DR15 that fulfill these criteria.

\subsection{ALMA}

The CO data used for the work presented here are taken from the ALMA-MaNGA QUEnching and STar formation (\aq) survey (Lin et al. in prep).  The ALMaQUEST survey consists of 46 galaxies at $z \sim 0$ selected from MaNGA with ALMA CO (1-0) follow-up observations obtained in Cycles 3, 5 and 6.  The \aq\ sample and data reduction are described in detail by Lin et al (in prep).  In brief, the sample includes both star forming and `green valley' galaxies selected from the MaNGA DR14 and DR15 with stellar masses in the range $10<$ \logms\ $<11.5$.  Since the \aq\ sample was designed to investigate galaxies with a wide range of star formation rates, the survey is not complete in terms of mass or volume, and the relative fractions of star forming, star bursting and green valley galaxies are not representative of the local universe.  However, within the stellar mass range sampled, the \aq\ survey allows us to study a broad range of star formation rates, and is hence well suited to studies that aim to investigate the physical mechanisms that both promote, and quench, star formation.

ALMA observations were obtained with proposals 2015.1.01225.S, 2017.1.01093.S, 2018.1.00558.S (PI Lin) and 2018.1.00541.S (PI Ellison) using the C43-2 configuration between 2016 and 2019.  In this compact configuration, the beam size of the ALMA observations is well matched to the point spread function of MaNGA observations (approximately 2.5 arcsecs) allowing the datasets to be matched to the same spatial sampling grid of 0.5 arcsec pixels (an angular scale that corresponds to 0.3 kpc at the median redshift of the sample, $z$=0.03). The largest angular scale of the ALMA observations is set to match the size of the MaNGA IFU.  For most of the galaxies in the \aq\ sample this choice is expected to capture most of the CO emission.  However, we note that the analysis in this paper is performed on images mapped to the same spaxel grid as MaNGA, such that we are using molecular gas measurements on the scale of 0.5 arcsec, rather than global measurements.  Missing flux is therefore not a concern for the analysis presented here. 

The CO luminosities are converted to \sigh2\ by assuming a constant $\alpha_{CO}$ = 4.3 M$_{\odot}$ pc$^{-2}$ (K km s$^{-1}$)$^{-1}$, e.g. Bolatto et al. (2013); the implications of adopting a fixed conversion factor are discussed below and in the Results section. As with other measures of surface density used in this work (e.g. \sigstar, \sigsfr), we correct \sigh2\ for inclination. In the analysis presented here, we require a S/N $\ge$ 3 in the integrated CO line flux, which mirrors the quality requirement of the optical emission lines.  There are 15,330 spaxels  in the \aq\ sample that fulfill both the MaNGA quality control criteria described in the previous sub-section and the CO S/N cut.

\subsection{Offsets from local scaling relations:  \dsigsfr, \dsfe\ and \dfgas}\label{offset_sec}

Several higher order spaxel quantities are computed from the MaNGA and ALMA gridded data, to be used as measures of offsets from resolved scaling relations.  These offset metrics allow us to compute how deviant spaxel properties are from the `norm'.  Specifically, we compute the following three offset quantities: \dsigsfr, \dsfe\ and \dfgas, all of which are computed in log space relative to the median of their control spaxels, such that for property $X$, we can define:

\begin{equation}
  \Delta X = \log X - <\log X_{\rm control}>_{\rm med}
\end{equation}

\dsigsfr\ is computed for every spaxel in the \aq\ sample which passes the quality control criteria described above.   First, a set of reference spaxels, used to define the rSFMS, is selected from all star-forming spaxels in the DR15 (not just those in the \aq\ sample)\footnote{We do not use the \aq\ sample itself to define the rSFMS, since the high fraction of green valley galaxies in the sample means that spaxels with low \sigsfr\ are over-represented and thus pull down the median value of the control pool at fixed \sigstar.  This would lead to a systematic reduction in \dsigsfr\ of $\sim$ 0.1 dex.  However, even if we had used the \aq\ sample to define the rSFMS, this would not have changed any of the conclusions in this paper, since the overall correlation strengths of \dsigsfr\ with gas fraction and SFE are robust against a systematic shift in \dsigsfr.}. For this reference set, we additionally require that the galaxy's inclination is $<$ 70 degrees (b/a$>$0.34), to avoid effects stemming from high inclinations.  To compute a \dsigsfr\ for any given spaxel, a set of `control' spaxels is drawn from the rSFMS reference sample, matched within 0.1 dex in \sigstar, 0.1 dex in total stellar mass  (of the host galaxy) and 0.1 in $R/R_e$, with typically several hundred control spaxels identified for each spaxel. Matching in $R/R_e$ mitigates radial dependences that are known to exist in, for example, SFR and O/H (e.g. Gonzalez-Delgado et al. 2014, 2015; Belfiore et al. 2017; Sanchez-Menguiano et al. 2018).  We do not match explicitly in O/H, since we will later investigate the correlation of metallicity with \dsigsfr.  However, we tested the impact of including O/H in the matching parameters and find a systematic increase in \dsigsfr\ of $\sim$0.01 dex, which is negligible. Although we do not expect \dsigsfr\ to depend strongly on total stellar mass (e.g. Ellison et al. 2018), other spaxel properties, such as O/H (which we will return to in Section \ref{discuss_sec}) do show a dependence on M${_\star}$ (e.g. Hwang et al. 2019).  We therefore include stellar mass matching for all of our offset metrics, for consistency.  In summary, \dsigsfr\ can be broadly thought of as the offset from the rSFMS (with additional controls for total stellar mass and position within the galaxy), and hence can identify starburst spaxels.  The process of computing a \dsigsfr\ from a sample of matched spaxels, rather than from an established fit to the rSFMS, avoids parametric uncertainties related to fitting approaches and observational effects such as resolution (e.g. Hani et al. 2019).

Next, we compute \dsfe, which captures the relative star formation efficiency of a given spaxel, compared to the norm.  To first order, \sigsfr\ is regulated by the surface density of neutral gas (e.g. Kennicutt 1998b).  However, combining spatially resolved maps of both HI and H$_2$ has revealed that it is the molecular gas that drives this relationship (Bigiel et al. 2008; Leroy et al. 2008; Schruba et al. 2011).  A first order estimate of \dsfe\ could therefore simply quantify the offset of a given spaxel's \sigsfr\ relative to the average value at fixed \sigh2.  However, further subtle complexities are also encoded into the Kennicutt-Schmidt relation, which shows  dependence on, for example, the total stellar mass of the galaxy and the galactocentric radius (e.g. Leroy et al. 2013;  Bolatto et al. 2017).   Multiple matching parameters should therefore be included in the computation of \dsfe.

Our approach to calculating \dsfe\ is analogous to our calculation of \dsigsfr, in that we compute the offset of the y-variable (\sigsfr\ for the calculation of both \dsigsfr\ and \dsfe) relative to the x-variable (\sigstar\ for the calculation of \dsigsfr\ and \sigh2\ for the calculation of \dsfe).  First, we first identify a pool of control spaxels with `normal' star formation rates to define a reference Kennicutt-Schmidt relation.  This Kennicutt-Schmidt reference set is comprised of spaxels from the \aq\ sample with $-0.5 <$ \dsigsfr\ $<0.5$ and galaxy b/a$>0.34$, to avoid inclination effects.   Similar to the process described above for \dsigsfr, \dsfe\ is then computed as the offset of a given spaxel's \sigsfr\ from those spaxels in the control sample that are matched within 0.1 dex in \sigh2, 0.1 dex in total stellar mass (of the host galaxy), 0.1 in $R/R_e$ and 0.1 dex in O/H.  These matching criteria will mitigate dependences of $\alpha_{CO}$ on metallicity and galactocentric radius (Narayanan et al. 2011; Bolatto, Wolfire \& Leroy 2013; Sandstrom et al. 2013; Cormier et al. 2018). \dsfe\ is then the offset of \sigsfr\ of a given spaxel relative the median \sigh2\ of its control spaxels, with positive values of \dsfe\ indicating greater SFEs (or shorter depletion times).  As emphasized for the calculation of \dsigsfr, an important characteristic of the above procedure for calculating \dsfe\ is that it does not rely on fitting the Kennicutt-Schmidt relation, the solution to which depends on both dataset specific quantities, such as SFR tracer (e.g. Leroy et al. 2013) and the range of SFRs included in the fit (e.g. Gao \& Solomon 2004), as well as the fitting method itself (e.g.  Lin et al. 2019).

The final offset metric used in the current paper is \dfgas, which captures the relative gas fraction of a given spaxel.  \dfgas\ is defined as the offset from the molecular gas main sequence (\sigstar\ -- \sigh2), which forms a third dimension with the resolved SFMS and the Kennicutt-Schmidt relation (Lin et al. 2019).  Specifically, \dfgas\ is computed as the offset of \sigh2\ at fixed \sigstar, with additional spaxel matching in total stellar mass of the host galaxy (within 0.1 dex), $R/R_e$ (within 0.1) and O/H (within 0.1 dex).  As for \dsfe, control spaxels are drawn from the combined ALMA plus MaNGA spaxel dataset with normal star formation rates, i.e. $-0.5 <$ \dsigsfr\ $<0.5$ and inclinations (b/a$>0.34$).  

Our conversion from CO luminosity to \sigh2\ assumes a fixed $\alpha_{CO}$ that is representative of a `normal' disk galaxy ($\alpha_{CO}$=4.3 M$_{\odot}$ pc$^{-2}$ (K km s$^{-1}$)$^{-1}$).  Starburst galaxies are known to exhibit lower values of $\alpha_{CO}$ (e.g. Downes \& Solomon 1998; Bryant \& Scoville 1999; Papadopoulos et a. 2012; Sargent et al. 2014), and resolved studies of gas and dust have indicated lower values of $\alpha_{CO}$ are also common at small galactic radii (Sandstrom et al. 2013; Cormier et al. 2018).  It is therefore likely that our assumption of a fixed $\alpha_{CO}$ is incorrect, with smaller values likely to be more appropriate for the inner starbursting spaxels of our sample (e.g. Magdis et al. 2012; Sargent et al. 2014).  Smaller values of $\alpha_{CO}$ will reduce the \sigh2\ in a given spaxel, shifting it leftwards on the \sigh2\ -- \sigsfr\ plane, and hence above the normal Kennicutt-Schmidt relation, i.e., the effect of variable $\alpha_{CO}$ in starburst spaxels would likely be to increase \dsfe\ and decrease \dfgas.  Although the analysis in the following sections uses only a fixed  $\alpha_{CO}$, we also discuss the qualitative effect of a variable conversion factor, given the probable dependence of \dsfe\ and \dfgas.

\subsection{Central starburst sample selection}\label{sample_sec}

The focus of the current paper is galaxies with central starbursts.   Although Ellison et al. (2018) showed that central starbursts are generally associated with galaxies located above the global SFMS, we select our central starburst sample based on local, rather than global properties.  Azimuthally averaged radial profiles of \dsigsfr\ are constructed for each galaxy in the \aq\ sample by taking median values within a sliding bin of width 0.5 $R/R_e$.  The bin centres are incremented by 0.1 $R/R_e$ in each iteration.  The bins are therefore not independent, but this approach allows us to represent more of the structure in the profile.  Radial profiles for all galaxies in the \aq\ sample were visually inspected to identify those with star formation rates within 0.5 $R_e$ elevated by at least 50 per cent.  In practice, this was implemented by requiring the median \dsigsfr\ within 0.5 $R_e$ exceeds 0.2 dex.  Whilst this is a somewhat subjective definition of central starburst, the exact definition does not affect our conclusions.  In Ellison et al. (2019) we present a more general assessment of the primary drive of \sigsfr, and its regulation above the rSFMS, for all spaxels in the \aq\ sample.

\begin{figure}
	\includegraphics[width=\columnwidth]{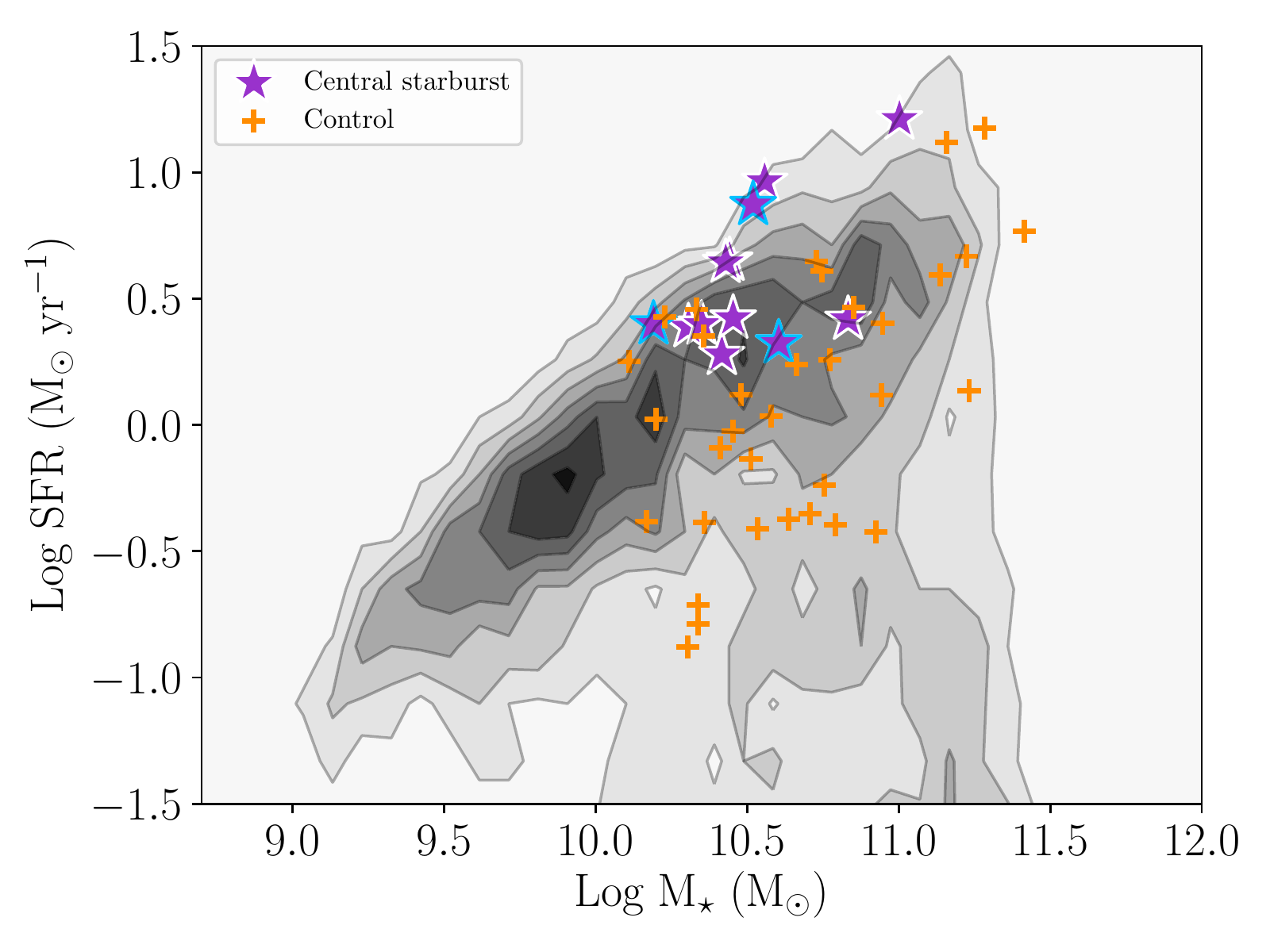}
        \caption{The distribution of total stellar masses and SFRs for the complete DR15 MaNGA sample is shown in grey scale.  Individual symbols indicate the 46 galaxies in the \aq\ sample, with the central starbursts shown with purple stars (outlined in light blue if they show features of morphological disturbance).   }
    \label{MS_fig}
\end{figure}

\begin{figure}
	\includegraphics[width=\columnwidth]{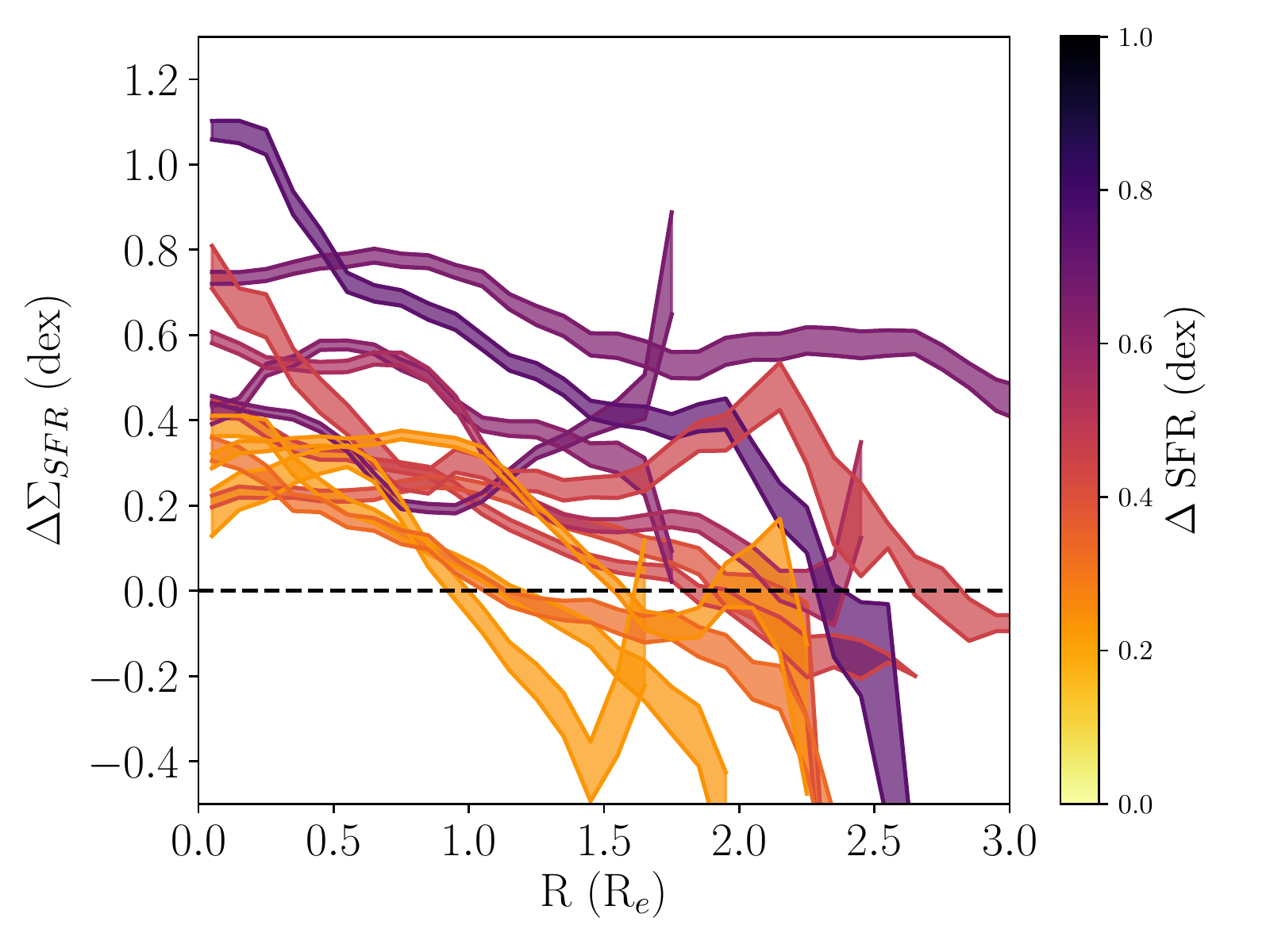}
        \caption{Azimuthally averaged radial profiles of \dsigsfr\ for the sample of central starburst galaxies identified within the \aq\ sample.  Curves are colour-coded by the offset of the galaxy's total SFR from the global star forming main sequence.  Profile widths in this and all subsequent figures indicate the RMS spaxel value in each radial bin, divided by the square root of the number of spaxels in that bin.}
    \label{SB_profiles_fig}
\end{figure}

12 central starburst galaxies are selected from the \aq\ sample; their total stellar masses and SFRs relative to the complete MaNGA DR15 are shown in Fig \ref{MS_fig} and their individual \dsigsfr\ radial profiles in Fig \ref{SB_profiles_fig}. Profiles are colour-coded by \dsfr, which is the offset from the global SFMS, computed at fixed total stellar mass and redshift.  As expected from the results of Ellison et al. (2018) the central starbursts all lie above the global SFMS, with a correlation between \dsfr\ and the strength of the central burst.

Fig. \ref{SB_profiles_fig} shows that, in most cases, the starburst galaxies in our sample are characterized by \sigsfr\ that is centrally enhanced, declining with increasing radius.  However, some of the galaxies in the central starburst sample have elevated \dsigsfr\ out to large radii.  The term `central starburst' should therefore not be taken to mean that star formation is boosted only in the central regions, but rather that \dsigsfr\ is enhanced at least in the centres.   Whilst two of the galaxies exhibit fairly modest central enhancements with median central \dsigsfr\ $\sim 0.2$ dex, the remaining 10 all show enhancesments of at least a factor of two.  One third of the sample shows central star formation enhancements of at least a factor of 3 and one galaxy is centrally boosted by over an order of magnitude above expectations from the rSFMS.  Overall, the magnitude of the central starbursts within our sample spans almost a factor of ten.    

\begin{figure*}
	\includegraphics[width=19cm]{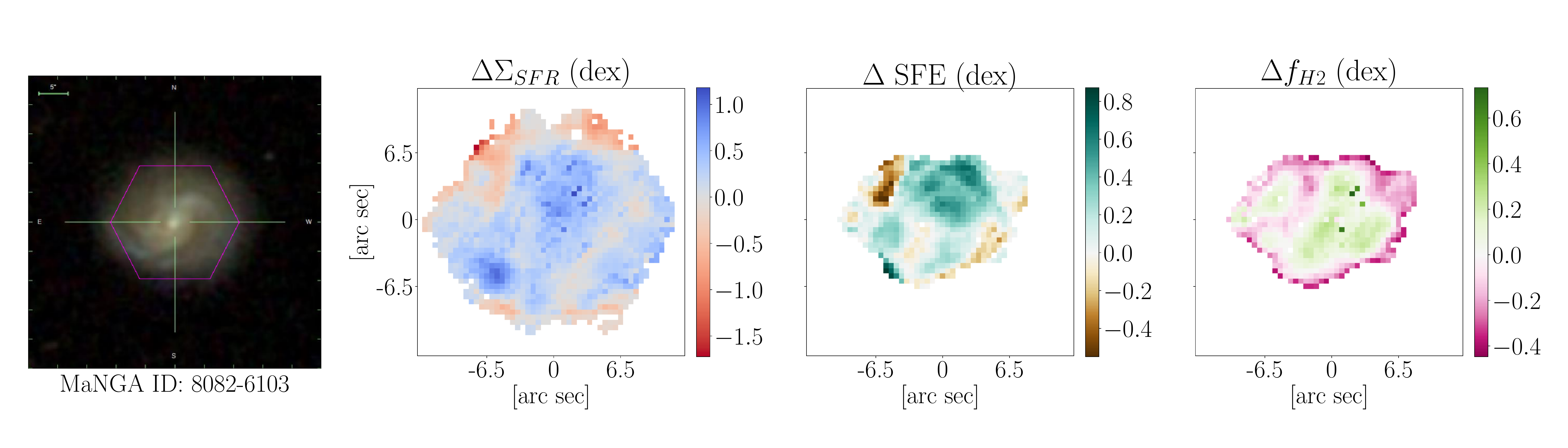}
	\includegraphics[width=19cm]{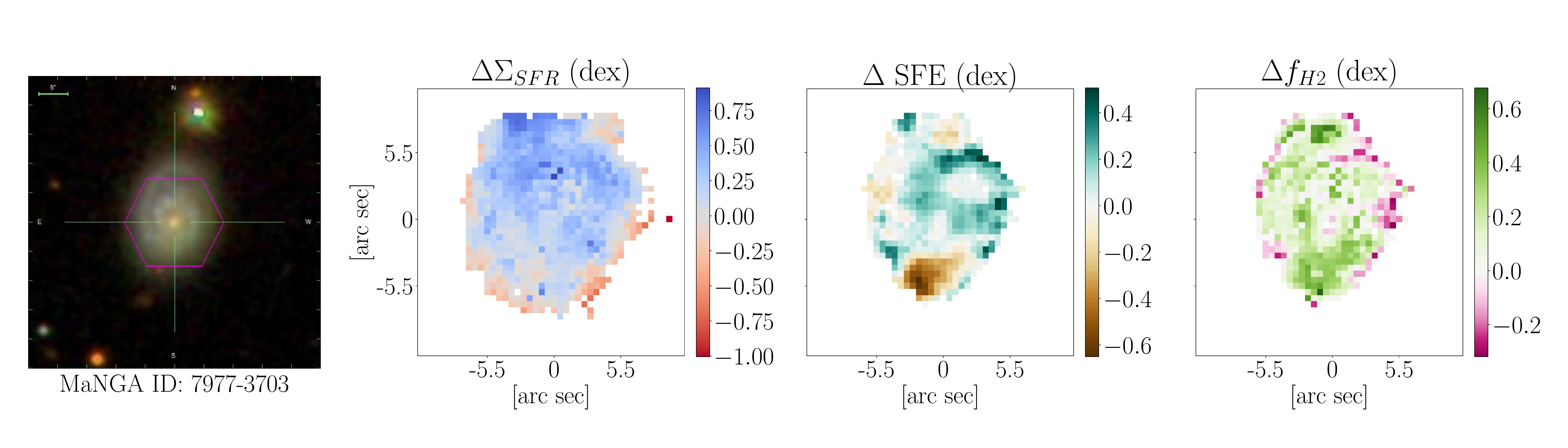}
	\includegraphics[width=19cm]{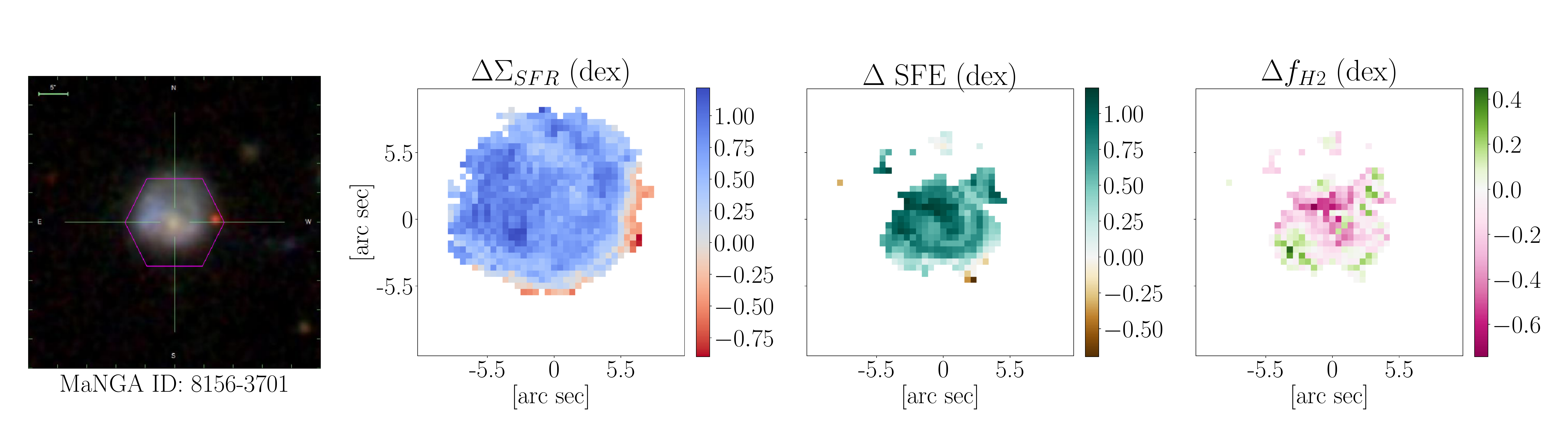}
	\includegraphics[width=19cm]{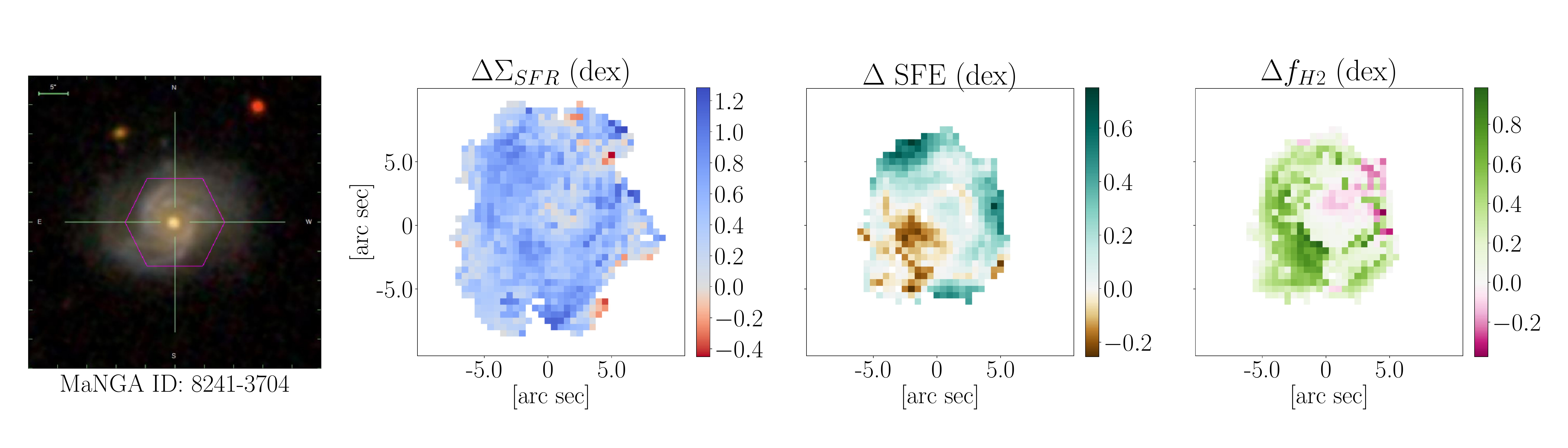}
        \caption{Maps of the \aq\ starburst sample.  The panels in each row, from left to right, show the SDSS $gri$ image with MaNGA IFU footprint overlaid in magenta and maps of \dsigsfr, \dsfe\ and \dfgas.  Divergent colour bars are used for the offset metrics, such that white indicates a normal value (of \sigsfr, SFE or gas fraction) and different colours (and shades thereof) indicate positive or negative offsets. }
    \label{maps_fig}
\end{figure*}

\begin{figure*}
        \addtocounter{figure}{-1}
	\includegraphics[width=19cm]{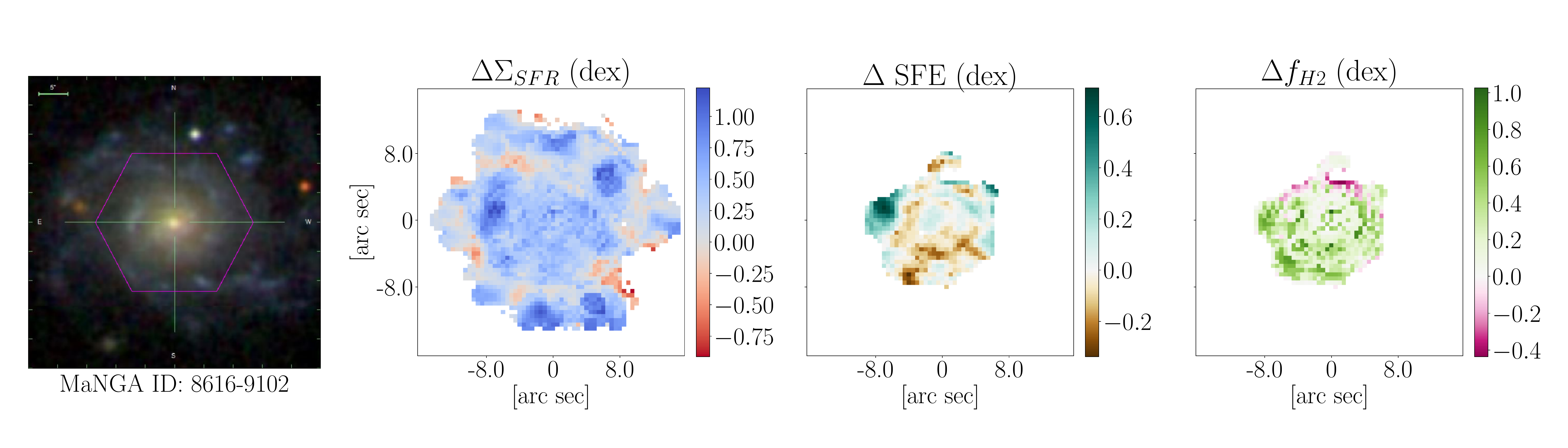}
	\includegraphics[width=19cm]{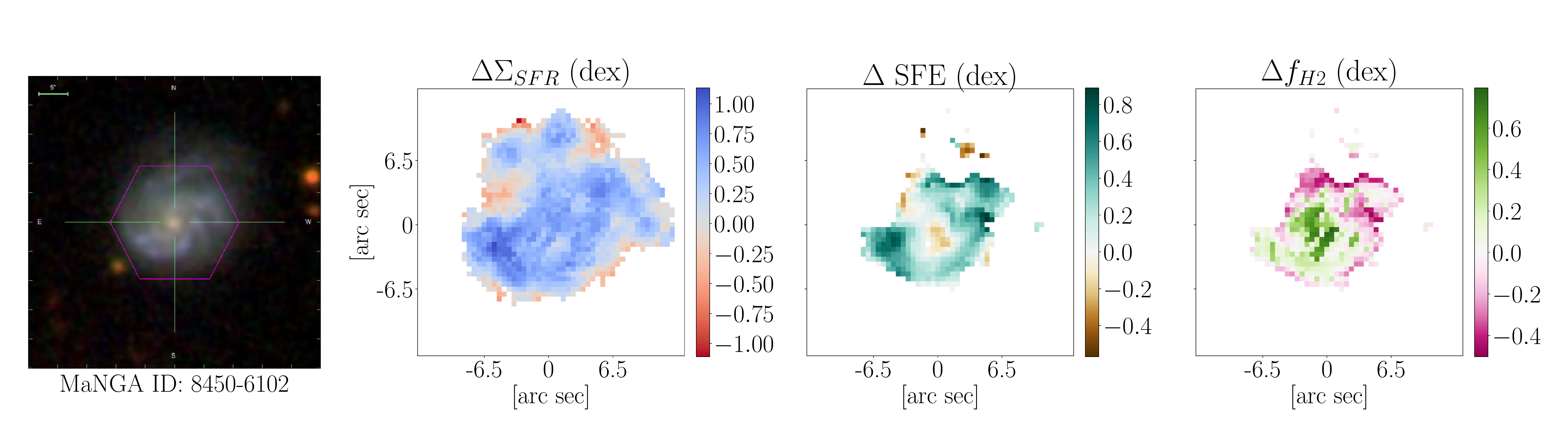}
	\includegraphics[width=19cm]{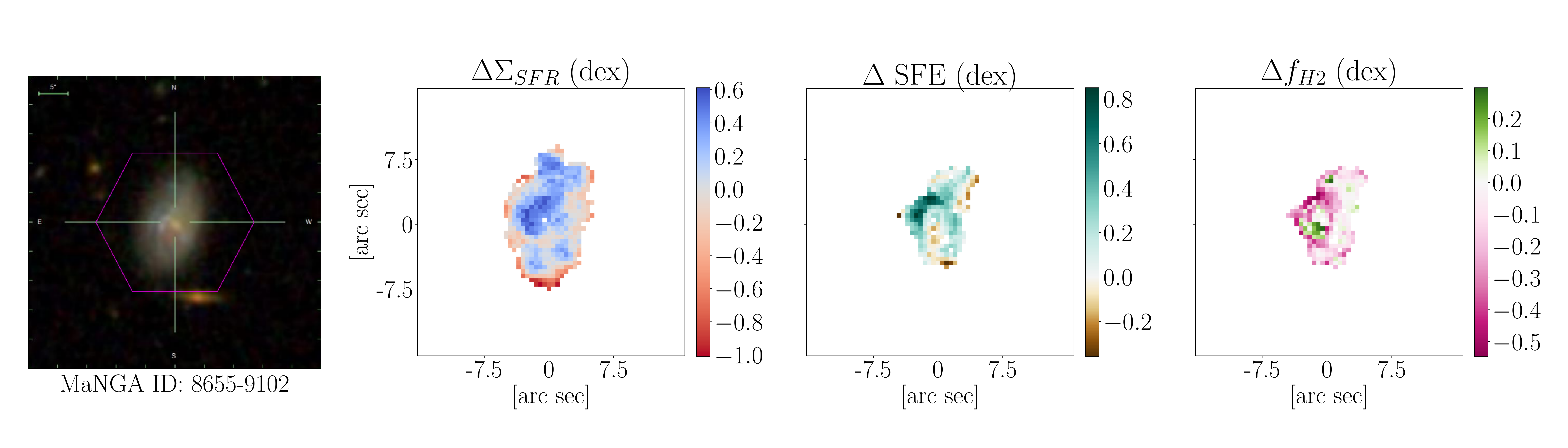} 
	\includegraphics[width=19cm]{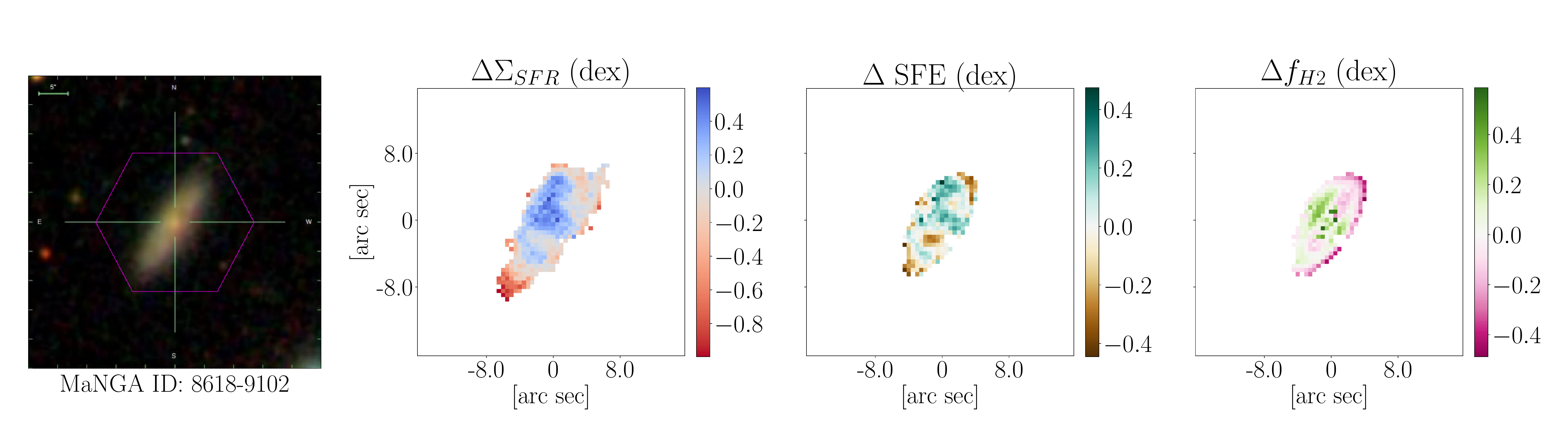}
        \caption{Maps  of the \aq\ starburst sample -- continued.}
\end{figure*}


\begin{figure*}
        \addtocounter{figure}{-1}
	\includegraphics[width=19cm]{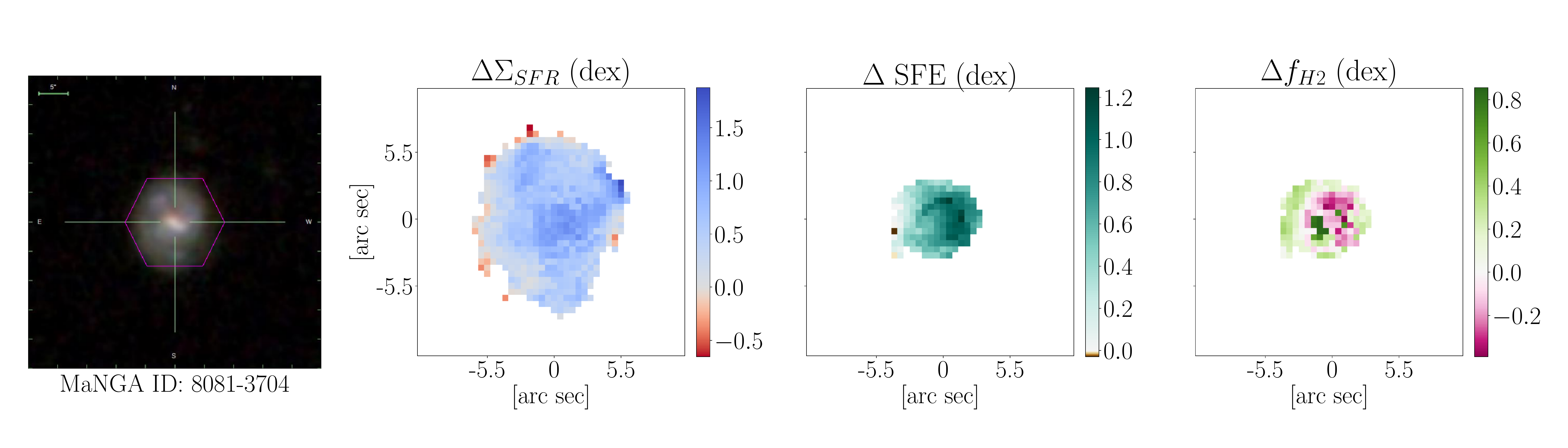}
	\includegraphics[width=19cm]{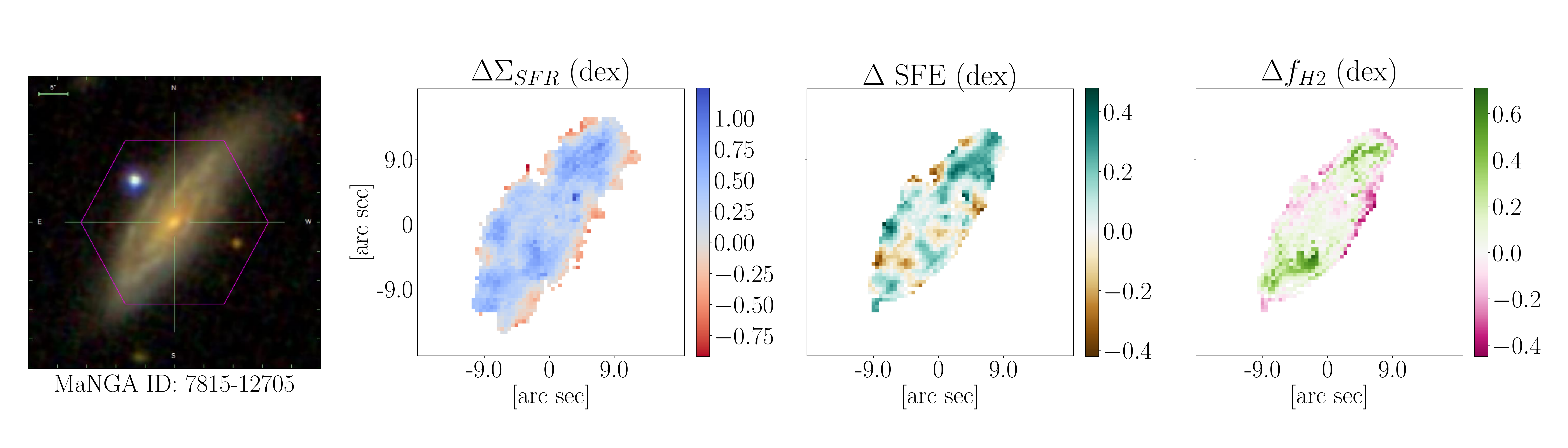}
	\includegraphics[width=19cm]{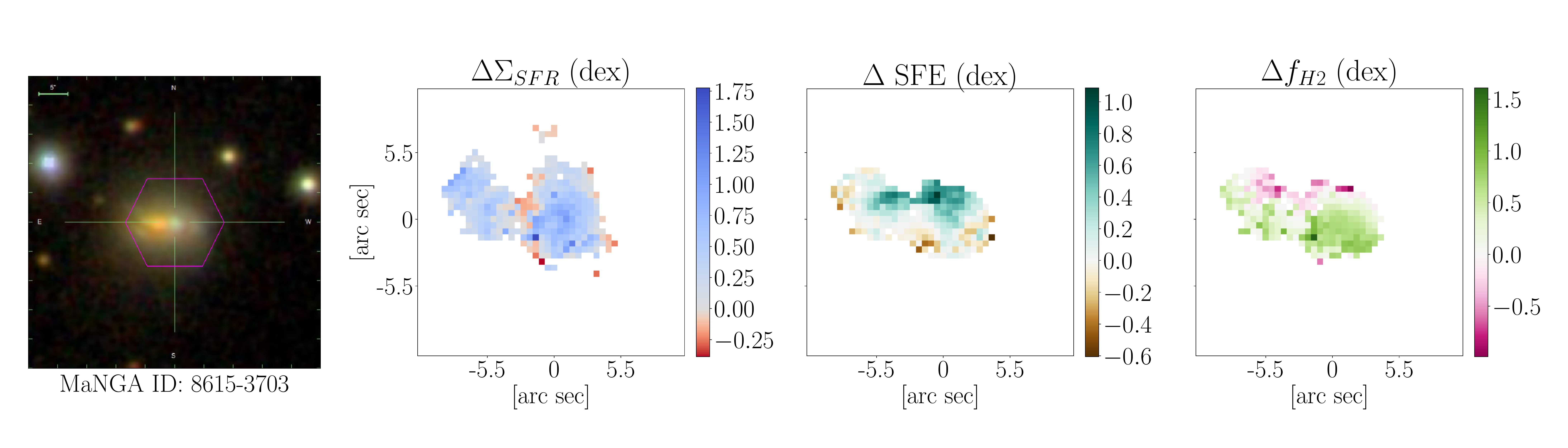}
	\includegraphics[width=19cm]{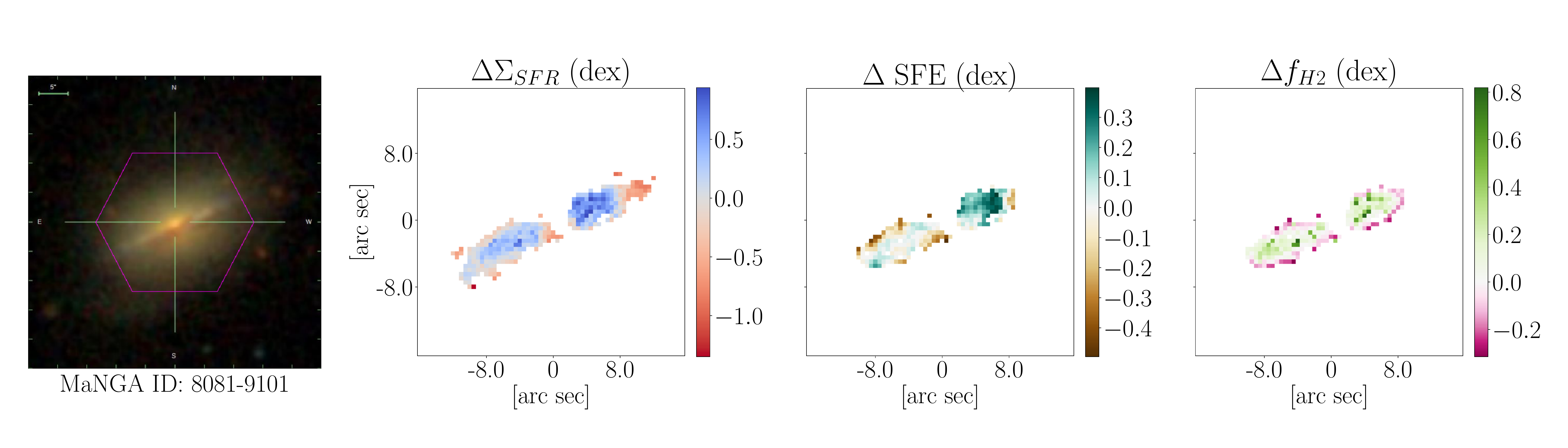} 
        \caption{Maps  of the \aq\ starburst sample -- continued.}
\end{figure*}


There are a total of 10,765 star-forming spaxels in the 12 central starburst galaxies that comprise our sample, of which 5278 spaxels also have \sigh2\ measurements, from which we can compute \dsfe\ and \dfgas.  In Fig. \ref{maps_fig} we show the SDSS $gri$ images of our central starburst sample, labeled with their MaNGA plate-IFU ID, and maps of the offset quantities (\dsigsfr, \dsfe\ and \dfgas) used in the analysis presented here.  We note that \dsigsfr\ values can be calculated for a larger number of spaxels in a given galaxy than \dsfe\ or \dfgas, since we only require spaxels to be robustly detected in the MaNGA data and be classified as star-forming.  Calculation of \dsfe\ and \dfgas\ has the additional requirement that the CO line is detected with S/N $>$ 3, hence the maps for these metrics may be less complete than for \dsigsfr.   More extensive spaxel maps of other kpc-scale properties are presented in the \aq\ survey paper (Lin et al. in prep).

The SDSS images shown in Fig. \ref{maps_fig} reveal that the majority of our central starburst sample are morphologically normal galaxies, with only three  (8156-3701, 8615-3703 and 8081-9101) showing signs of merger features.  These three galaxies are outlined in light blue in Fig. \ref{MS_fig}. Mergers are a commonly cited mechanism for inducing central starbursts, often proposed to be responsible for populating a distinct sequence on the Kennicutt-Schmidt (\sigh2\ -- \sigsfr) plane (e.g. Daddi et al. 2010; Genzel et al. 2010).  Whilst mergers do indeed trigger starbursts (e.g. Ellison et al. 2008b; Scudder et al. 2012) that are centrally concentrated (Ellison et al. 2013; Thorp et al. 2019; Pan et al. 2019), our sample demonstrates that while (some) mergers can trigger starbursts, other mechanisms can also do so (see also Ellison et al. 2018).  Since there are other mergers in the \aq\ sample that are not included in the starburst sample presented here, we reserve a detailed analysis of interacting galaxies for a future paper.

The maps in Fig. \ref{maps_fig} demonstrate that azimuthally averaged radial profiles, such as those shown in Fig. \ref{SB_profiles_fig}, compress much of the spatial information.  A similar issue has been noted by Thorp et al. (2019) in their MaNGA analysis of post-merger galaxies.  We therefore analyse the gas properties of our central starburst sample using both spatially averaged and individual spaxel correlations.

\section{Results}\label{results_sec}

The main question posed in the current paper is: what drives the enhanced star formation rates (as quantified by \dsigsfr) in the centres of the 12 galaxies shown in Fig \ref{SB_profiles_fig}?  We consider two broad physical scenarios: 1) starbursts are driven by elevated SFEs;  2) starbursts are driven by elevated gas fractions.  We tackle this question by analysing the starburst sample in two principal ways.  First, we consider galaxy-by-galaxy properties, in which we spatially average either the central region, or compute azimuthally averaged radial profiles.  This approach has the benefit of reducing the complexity of the data (e.g. each galaxy can be represented as a single data point) and allows us to study galaxy-to-galaxy variations, to assess scatter and outliers (e.g. Vulcani et al. 2019a).  Whilst simplifying the data for visualization purposes, the spatial averaging erases the significant internal structure of the properties we are interested in (e.g. Fig. \ref{maps_fig}).  To complement this first approach, we therefore also analyze the ensemble spaxel properties of the entire sample, without regard to the galaxy of origin.  

\subsection{Results from spatially averaged profiles}

\begin{figure}
	\includegraphics[width=\columnwidth]{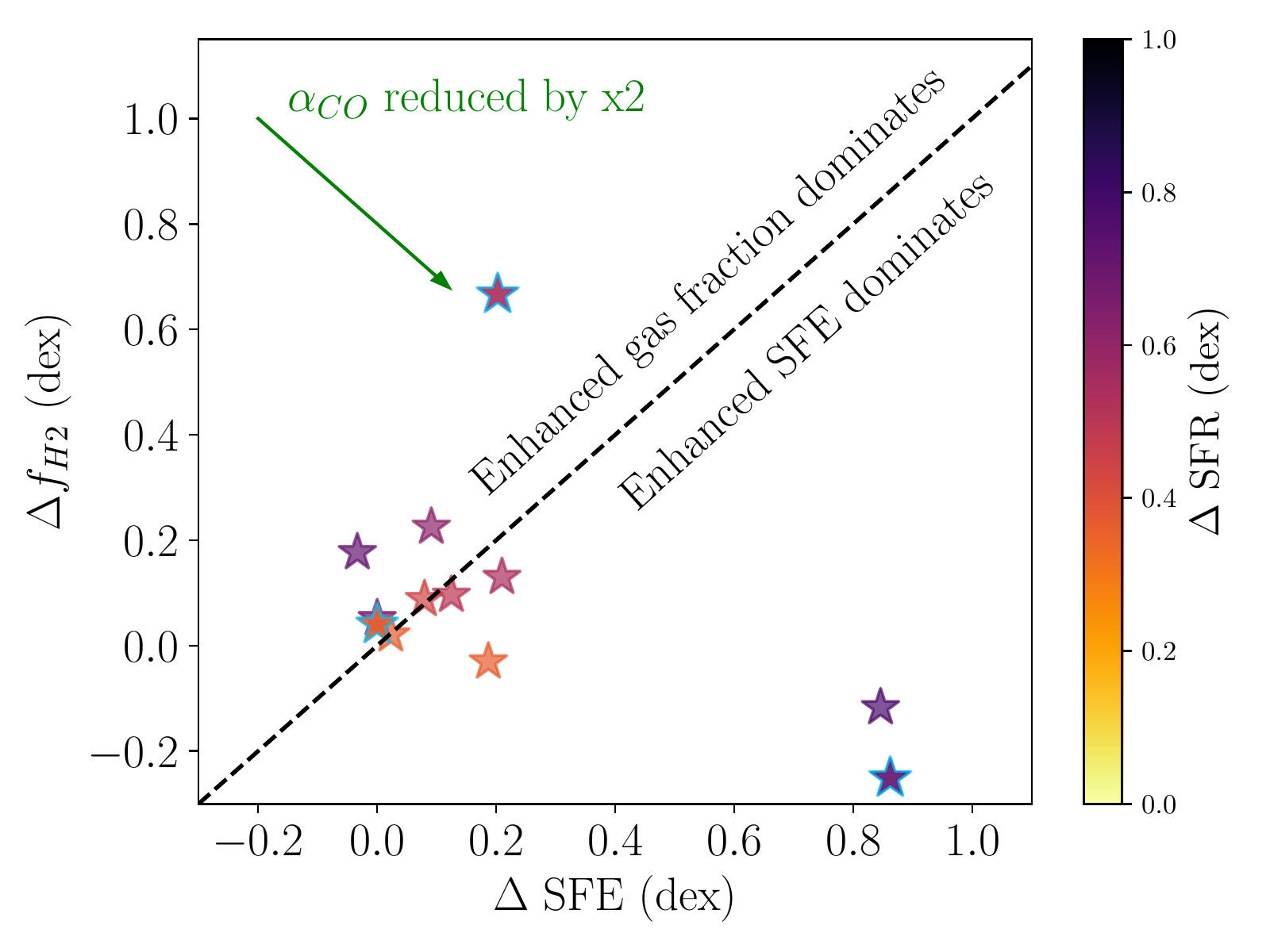}
        \caption{The median \dsfe\ and \dfgas\ within 0.5 $R_e$ for each of the 12 starburst galaxies is shown.  Points are colour-coded by the offset of the galaxy's total SFR from the global star forming main sequence; those edged in light blue indicate galaxy mergers. The dashed line divides the regimes in which gas fraction or SFE enhancements dominate. The arrow in the top left of the plot shows the correction if the conversion factor for the central regions of starbursts was reduced by a factor of two.}
    \label{central_fig}
\end{figure}

\begin{figure*}
	\includegraphics[width=\columnwidth]{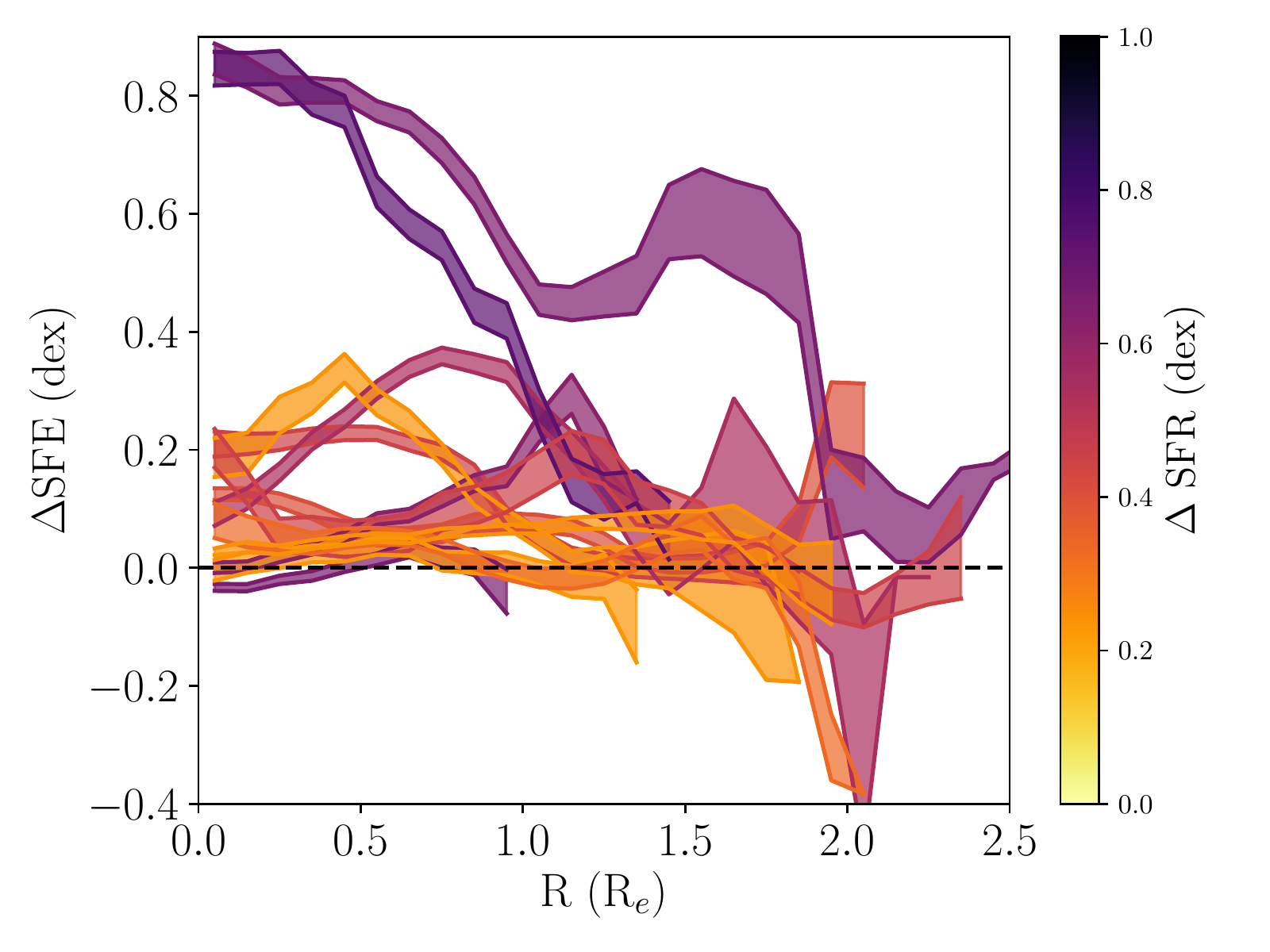}
 	\includegraphics[width=\columnwidth]{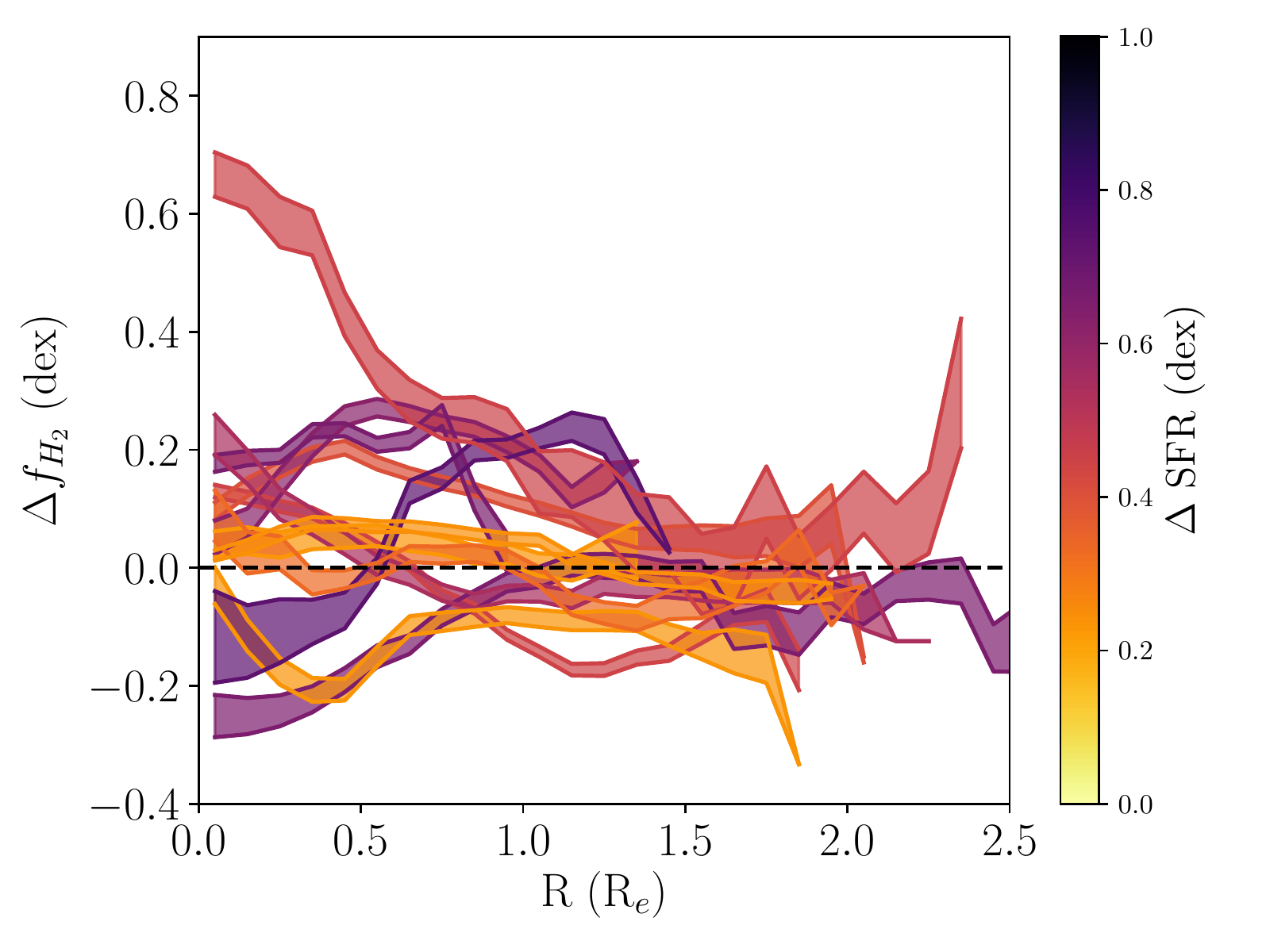}
       \caption{Azimuthally averaged radial profiles of \dsfe\ (left panel) and \dfgas\ (right panel) for the sample of central starburst galaxies identified within the \aq\ sample. Curves are colour-coded by the offset of the galaxy's total SFR from the global star forming main sequence.}
    \label{dsfe_dfgas_fig}
\end{figure*}

We begin our analysis by showing in Fig. \ref{central_fig} the median \dfgas\ and median \dsfe\ within 0.5 $R_e$ for our starburst sample.  Points are once again colour-coded by their global offset from the star forming main sequence and points representing the three galaxy mergers are edged in light blue.  The dashed line distinguishes the regimes in which enhancements in either gas fraction or SFE dominate.

Fig. \ref{central_fig} shows that, under our assumption of a fixed $\alpha_{CO}$, equal numbers of galaxies are dominated by boosts in gas fraction or SFE.  At face value then, the distribution of points in Fig. \ref{central_fig} could be interpreted as evidence that both elevated gas fractions and enhanced SFEs play a role in the starburst, with different mechanisms dominating in different galaxies.  However, recall that we have assumed a Milky Way-like conversion factor for all spaxels in our sample, with value of $\alpha_{CO}$ = 4.3 M$_{\odot}$ pc$^{-2}$ (K km s$^{-1}$)$^{-1}$.  As described in Section \ref{offset_sec}, a variable $\alpha_{CO}$, whose value will likely be lower for starbursting spaxels (and those at small radii), acts to shift points from the gas fraction dominated regime into the SFE dominated regime, by simultaneously increasing \dsfe\ and reducing \dfgas.

The most extreme values of $\alpha_{CO}$ are found in ultra-luminous infrared galaxies (ULIRGs) where values are typically $\alpha_{CO} \sim 0.8$ M$_{\odot}$ pc$^{-2}$ (K km s$^{-1}$)$^{-1}$ (e.g. Downes \& Solomon 1998).  Adopting such a low conversion factor would reduce \sigh2\ by a factor of $\sim$ 5 and hence alter both \dfgas\  and \dsfe\ by $\sim$ 0.7 dex. However, the central starburst galaxes studied here are not ULIRGs and have SFR enhancements that are much more modest.  Therefore, whilst the adopted value of $\alpha_{CO} = 4.3$ M$_{\odot}$ pc$^{-2}$ (K km s$^{-1}$)$^{-1}$ is likely to be an over-estimate, observational evidence of modestly starbursting galaxies, like those studied here, indicate that the correction required to $\alpha_{CO}$ will not be large.  For example, based on a comparison of CO and dust based molecular gas masses, Genzel et al. (2015) found very little dependence on $\alpha_{CO}$ for galaxies less than 0.6 dex above the global SFMS.  Sargent et al. (2014) present a model for conversion factors that depends in a continuous way on (global) SFR enhancements above the main sequence (see also Magdis et al. 2012).  For galaxies that lie above the global SFMS by a factor of a few (which is typical of our sample, see Fig. \ref{MS_fig} and colour coding in Fig. \ref{central_fig}), $\alpha_{CO}$ is reduced by a factor of 2--3, in turn reducing \dfgas\ by 0.3--0.5 dex.  Whilst we do not attempt to estimate corrected conversion factors for individual spaxels, for illustration purposes, the vector in the top left of Fig. \ref{central_fig} shows the affect of reducing $\alpha_{CO}$ by a factor of two for these central starburst galaxies.  A correction of this magnitude would be sufficient to move all of the data points currently in the enhanced gas fraction regime into the SFE dominated region of the figure.   Therefore, whilst the face-value conclusion from Fig. \ref{central_fig} is that gas fraction and SFE can both contribute to generating a central starburst, our assumption of a fixed $\alpha_{CO}$ may over-estimate the role of gas fraction from these centrally averaged values. Indeed, if $\alpha_{CO}$ is lower by a factor of two from the Milky Way value we have adopted, we would conclude from  Fig. \ref{central_fig} that enhanced SFEs universally dominated over the impact of gas fractions in the central starbursts.

As expected from Fig. \ref{MS_fig}, the colour coding in Fig \ref{central_fig} shows that our central starburst sample galaxies lie on or above the global star forming main sequence. Whilst our sample is small (and uncertainties in $\alpha_{CO}$ make the precise position of points in Fig. \ref{central_fig} uncertain) we note that the two galaxies with the greatest dominence of SFE over gas fraction offsets (in the lower right corner of Fig. \ref{central_fig}) have the highest offsets above the global main sequence.  This finding is consistent with that Sargent et al. (2014) who found that galaxies above the main sequence have elevated SFEs.

\begin{figure*}
	\includegraphics[width=19cm]{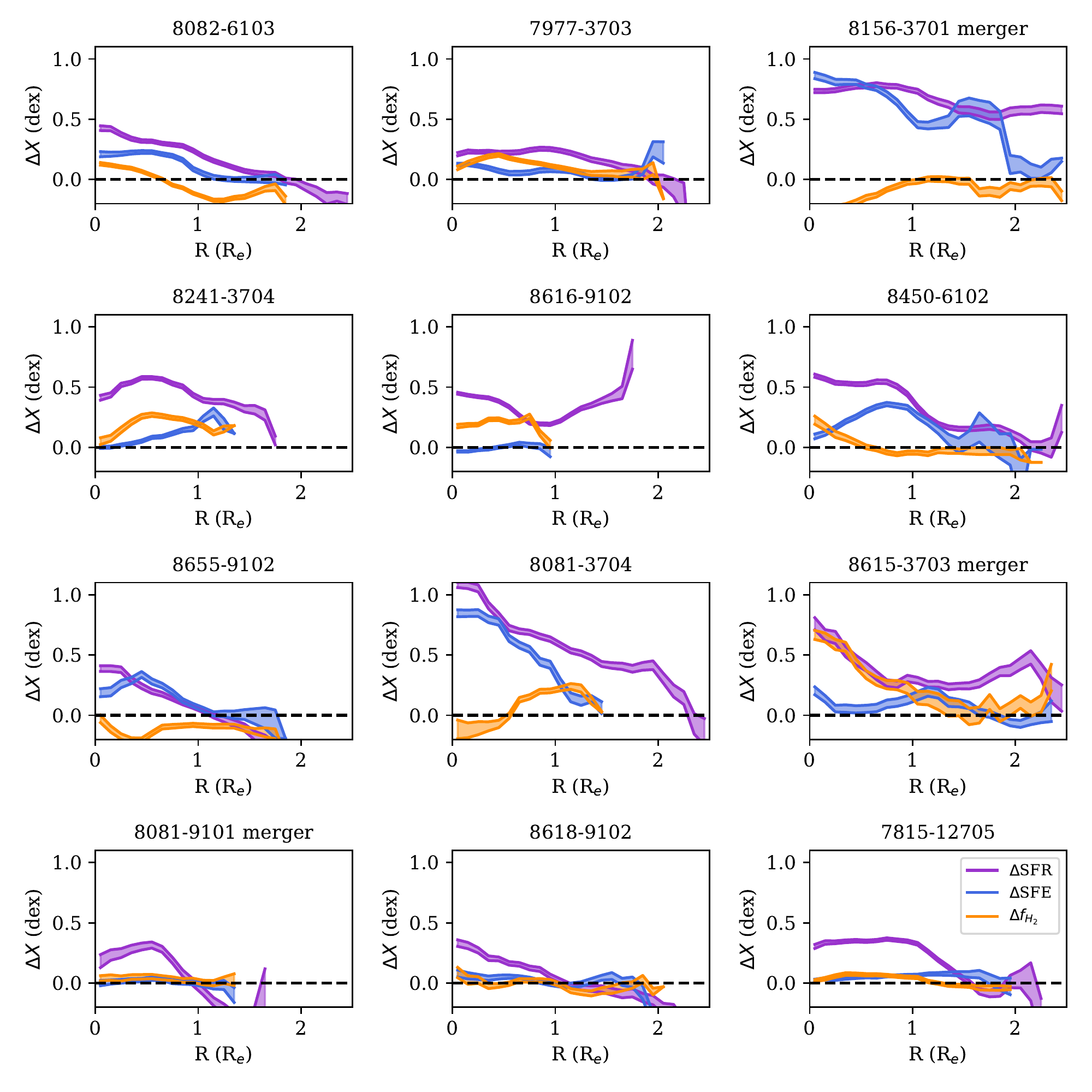}
        \caption{Azimuthally averaged radial profiles of \dsigsfr, \dsfe\ and \dfgas\ for the sample of central starburst galaxies identified within the \aq\ sample.  Purple, blue and orange curves show enhancements in star formation rate (\dsigsfr), star formation efficiency (\dsfe) and gas fraction (\dfgas) respectively.  Panel labels indidate the MaNGA plate-IFU number.}
    \label{all_profiles_fig}
\end{figure*}

Fig. \ref{central_fig} minimizes the dimensionality of our data, such that each galaxy can be represented by a single data point.  Whilst this has the advantage of a simple visual presentation, the significant spatial variations within a given galaxy (Fig. \ref{maps_fig}) are lost.  We therefore increase the spatial information captured in our analysis by showing in Fig. \ref{dsfe_dfgas_fig}  azimuthally averaged profiles of \dsfe\ (left panel) and \dfgas\ (right panel) for the central starburst galaxies in our sample.  As for the \dsigsfr\ profiles shown in Fig. \ref{SB_profiles_fig}, we again use a sliding bin of width 0.1 $R/R_e$ to generate the \dsfe\ and \dfgas\ profiles and colour code by global \dsfr.  

The \dsfe\ profiles in the left panel of Fig. \ref{dsfe_dfgas_fig} show a range of amplitudes and shapes.  For some of the galaxies, the profiles are fairly flat and have relatively normal SFEs (i.e. \dsfe\ $\sim$ 0), modulo our assumption of a fixed $\alpha_{CO}$ which could under-estimate these values.  Other galaxies show a negative gradient in \dsfe, with some galaxies exhibiting enhancements out to beyond an effective radius.  One of the two galaxies with the largest \dsfe\ enhancement is 8156-3701, which is one of the mergers in our sample.  As expected from the lower dimensional data in Fig. \ref{central_fig}, the two strongest central starbursts (with highest \dsfr) have the highest central \dsfe.

The right panel of Fig. \ref{dsfe_dfgas_fig} also shows a wide range of profile amplitudes for \dfgas, although 1/3 of the sample show approximately normal, or even surpressed, gas fractions out to 1 $R_e$.   There is no obvious connection between \dfgas\ and \dsfr, and the galaxies with the highest global \dsfr\ have fairly normal (within a factor of two) central gas fractions.   We remind the reader that a variable  $\alpha_{CO}$ will lead to lower limits on \dsfe\ and upper limits on \dfgas.  If we again suppose that $\alpha_{CO}$ may be reduced by a factor of two (0.3 dex) in the centres of these starburst galaxies, the majority of the (already modest) gas fraction enhancements seen in the right panel of Fig. \ref{dsfe_dfgas_fig} would disappear, and the \dsfe\ enhancements in the left panel Fig. \ref{dsfe_dfgas_fig} would be further enhanced.   The only galaxy in the sample that would still exhibit an elevated central gas fraction after applying a reduced $\alpha_{CO}$ is 8615-3703, another of the mergers in our sample.  We conclude that the evidence for enhanced SFEs in our starburst sample is robust to assumptions on the conversion factor.  There may also be a contribution from enhanced gas fractions, but this conclusion is dependent on the choice of $\alpha_{CO}$.  

Given the large scatter in the profiles of \dsfe\ and \dfgas\ in Fig.  \ref{dsfe_dfgas_fig}, and to investigate the possible interplay between the various offset metrics, it is also useful to inspect the radial profiles on a galaxy-by-galaxy basis.  In Fig. \ref{all_profiles_fig} we show the profiles of \dsfr, \dsfe\ and \dfgas\ for each of the 12 galaxies in our central starburst sample.  Purple, blue and orange curves show enhancements in star formation rate (\dsigsfr), star formation efficiency (\dsfe) and gas fraction (\dfgas) respectively.   Recall (e.g. Fig. \ref{maps_fig}) that \dsigsfr\ is computed for all star-forming spaxels, whereas a robust CO detection is additionally required to calculate \dsfe\ and \dfgas, leading to different radial extents for these quantities in some galaxies.

As may be expected from the results from the central averages shown in Fig.  \ref{central_fig}, Fig. \ref{all_profiles_fig} shows that some galaxies have \dsfe\ profiles that are higher than their \dfgas\ profiles, but other galaxies have \dfgas\ that exceeds \dsfe\ in all radial bins.   We emphasize again that enhancements in both SFE and gas fraction can exist out to large radii, at least 1.5 $R/R_e$ in many cases.  There is no obvious pattern in Fig. \ref{all_profiles_fig} that reveals whether gas fraction or SFE dominates in a given galaxy, or if the two metrics are correlated.

\subsection{Results from spaxel correlations}

In the previous subsection we argued that there is strong evidence, that is robust to possible changes in the CO conversion factor, for enhanced SFEs in our sample of 12 central starburst galaxies.  Some galaxies also show enhanced gas fractions, but these should be considered upper limits under our assumption of a Milky Way-like $\alpha_{CO}$.

However, reviewing again Fig. \ref{maps_fig}, it can be seen that significant azimuthal, as well as radial, structure exists in our spaxel maps.  Although we have referred to our sample as `central' starbursts, it is clear that the starbursts are clumpy, asymmetrically distributed and often extended.  We therefore conclude that azimuthally averaged profiles are likely to be insufficient to capture the parameters that are driving the starburst.  In this subsection, we therefore move beyond a galaxy-by-galaxy analysis, by investigating the properties of individual spaxels.

In Fig. \ref{correlation_fig} we quantify, for all 5278 spaxels in our starburst sample, the correlation between \dsigsfr\ and \dsfe\ (upper panel) and \dfgas\ (lower panel).  This is a direct test of whether the star formation rate surface density in any given kpc region (in our sample of 12 starburst galaxies) is regulated more strongly by SFE or gas fraction.  Most of the spaxels in our sample have positive values of \dsigsfr.  This is a direct result of our starburst selection (see Ellison et al. 2019 for a more general analysis of \dsigsfr\ in the \aq\ sample).  Fig. \ref{correlation_fig} shows that there is a correlation between \dsigsfr\ and both \dsfe\ (upper panel) and \dfgas\ (lower panel).  However, the correlation between \dsigsfr\ and \dsfe\ is both steeper, tighter and more significant (based on a Pearson correlation test) than the relation with \dfgas.  The correlation between \dsigsfr\ and \dsfe\ would be even steeper for a variable conversion factor, since \dsfe\ would systematically increase for increasing \dsigsfr.  Conversely, the correlation between \dfgas\ and \dsigsfr\ would become flatter if a variable $\alpha_{CO}$ were adopted.  Fig. \ref{correlation_fig} is therefore compelling evidence that enhanced SFEs play a dominant role in kpc-scale starburst process.

The upper panel of Fig. \ref{correlation_fig} suggests that at the highest \dsigsfr, spaxels exhibit particularly elevated star formation efficiencies, as indicated by the cloud of points above the dashed line with \dsfe\ $>$ 0.7.  These points come largely from just two of the galaxies in our sample: 8156-3701 and 8081-3704, which both exhibit the highest \dsigsfr\ and \dsfe\ profiles in our sample (see Fig. \ref{all_profiles_fig}).  With total stellar masses  \logms\ $\sim$ 10.5 and total log SFR $\sim$ 0.9 $M_{\odot}$ yr$^{-1}$, they also lie $\sim$ 0.7 dex above the global SFMS (see Fig. \ref{MS_fig}), the two most significant outliers from the SFMS in our sample. These findings again support those of Sargent et al. (2014) that galaxies wel above the global SFMS are primarily driven there by elevated SFEs.

\section{Discussion}\label{discuss_sec}

The primary conclusion of the work presented here is that starbursts are primarily driven by enhanced SFEs.  Due to the clumpy, asymmetric and spatially extended nature of starbursting regions, considering the ensemble of spaxel properties provides insight beyond galaxy-by-galaxy averages.  

\begin{figure}
	\includegraphics[width=\columnwidth]{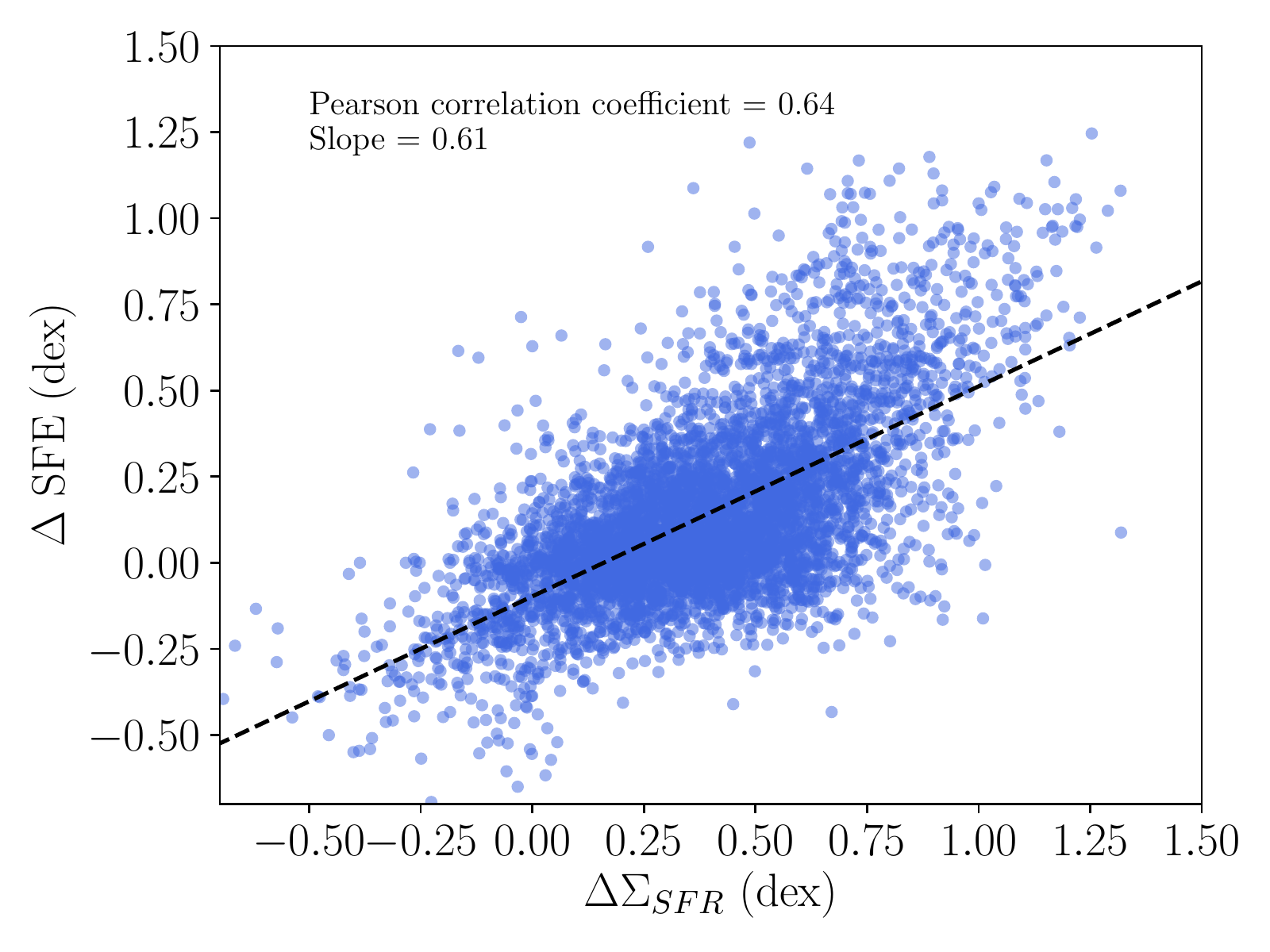}
	\includegraphics[width=\columnwidth]{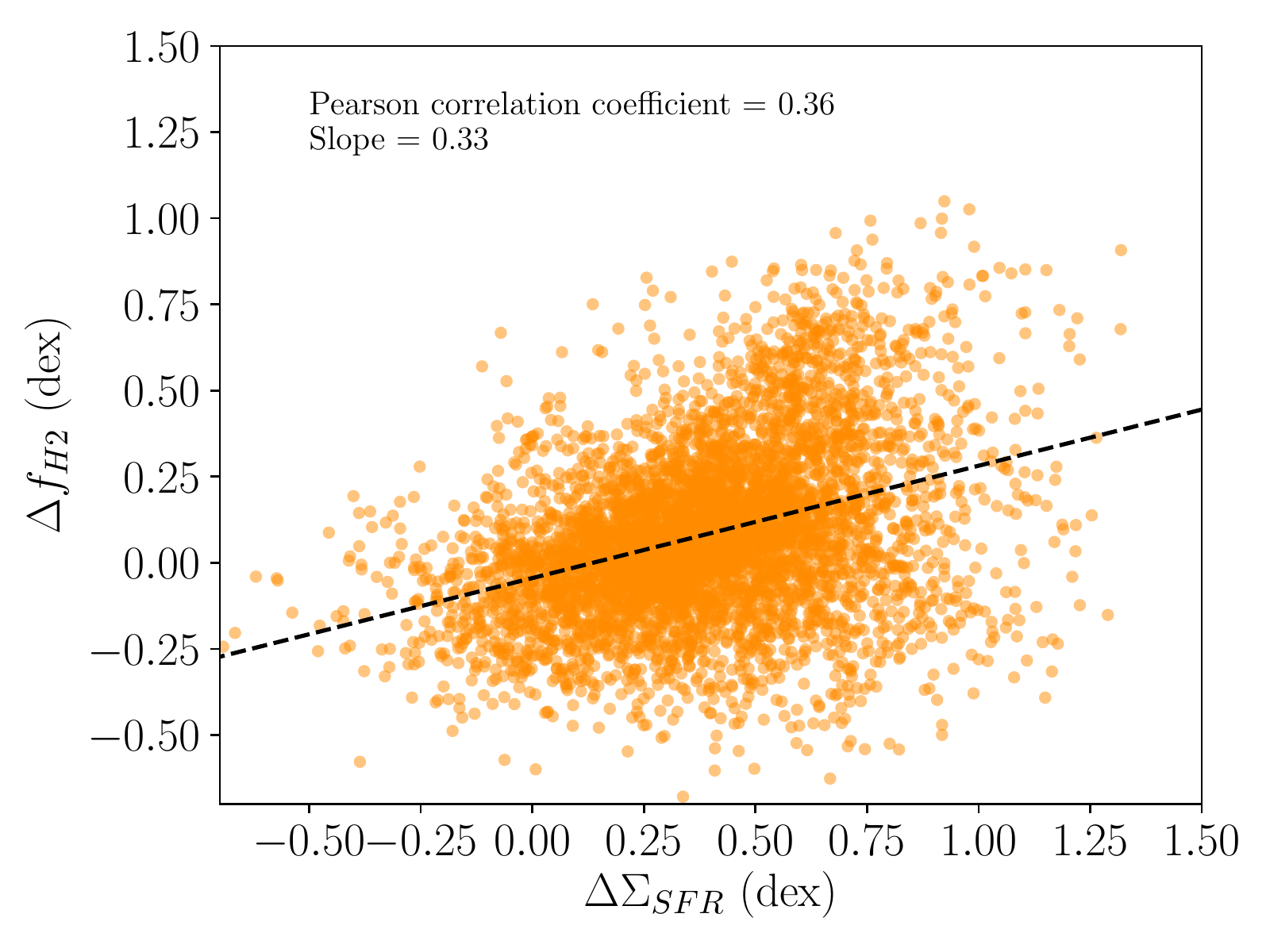}
        \caption{Is elevated gas fraction or enhanced SFE the primary driver of starbursts?  Upper panel: \dsigsfr\ vs. \dsfe.  Lower panel: \dsigsfr\ vs. \dfgas.  Dashed lines indicate a least squares fit. The stronger correlation between \dsigsfr\ and \dsfe\ (upper panel) than between \dsigsfr\ and \dfgas\ (lower panel) indicates that starbursts more strongly linked to enhanced SFEs than elevated gas fractions.}
    \label{correlation_fig}
\end{figure}

\begin{figure*}
	\includegraphics[width=19cm]{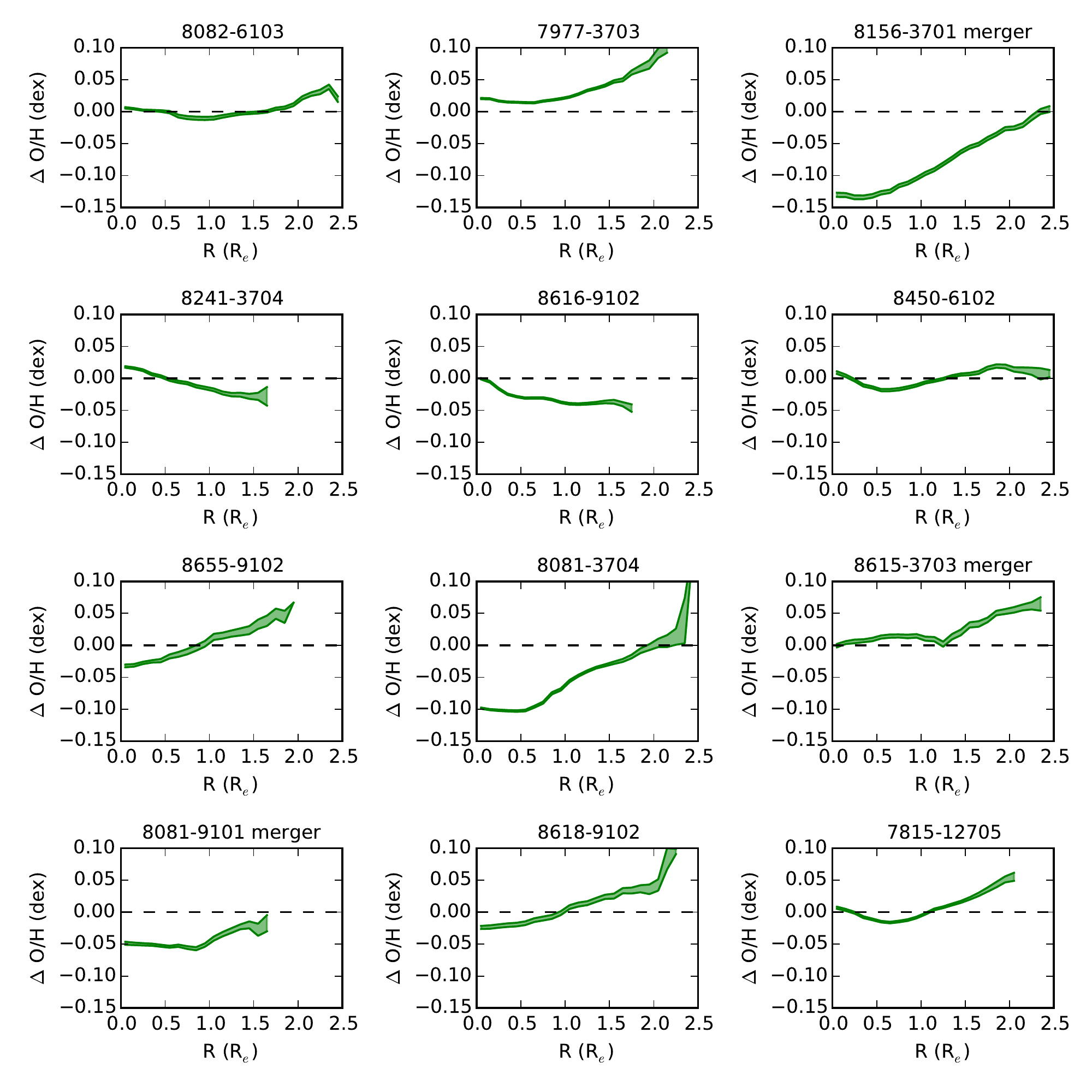}
        \caption{\doh\ profiles for the 12 central starburst galaxies in our sample.  Panels are in the same order as in Fig. \ref{all_profiles_fig}. }
    \label{doh_profiles_fig}
\end{figure*}

\begin{figure}
	\includegraphics[width=\columnwidth]{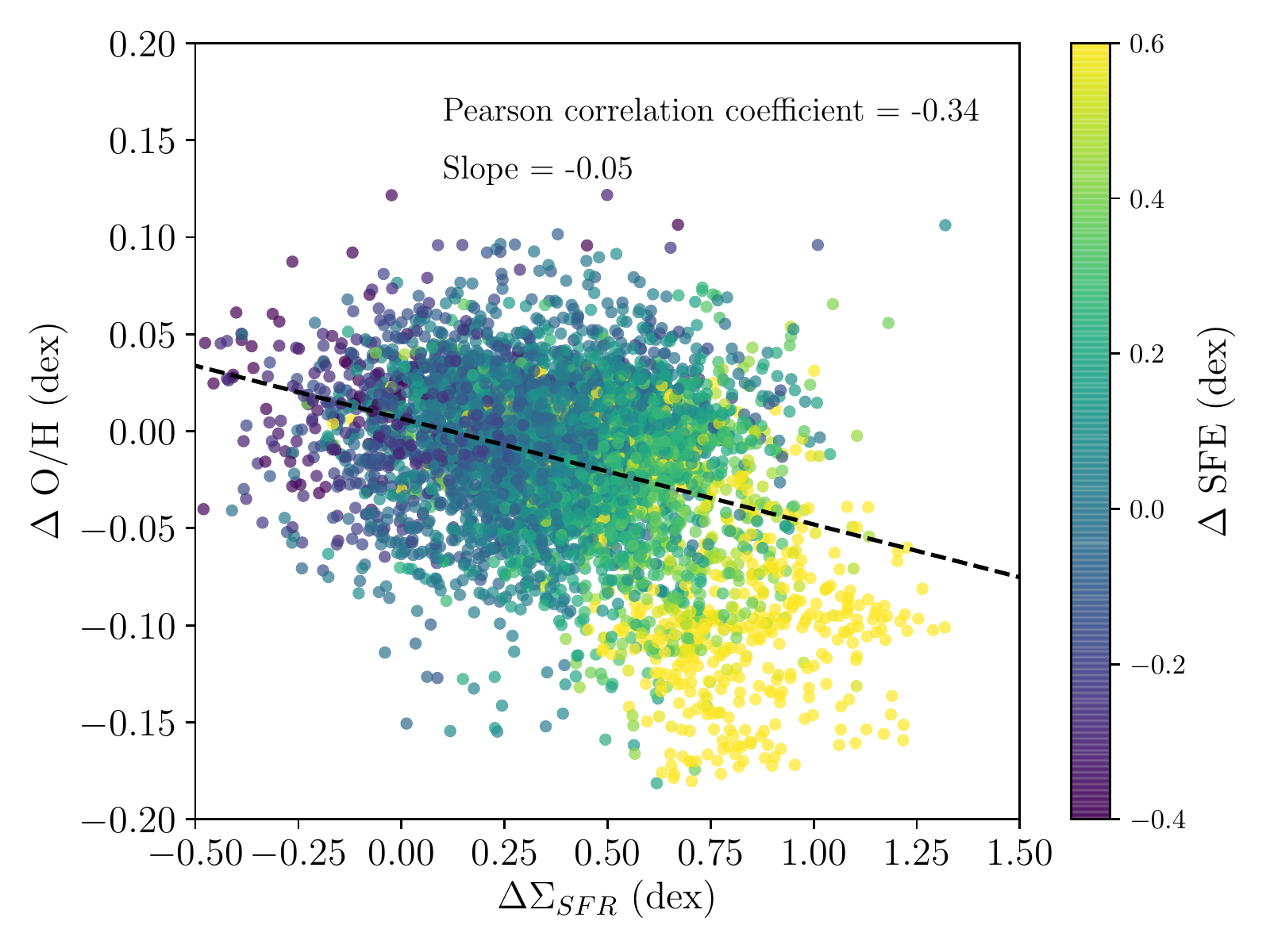}
        \caption{A weak anti-correlation between $\Delta$ O/H and \dsigsfr\ for $\sim$ 5000 spaxels in the \aq\ starburst sample indicates that starbursting regions are slightly metal-poor.}
    \label{doh_fig}
\end{figure}


Although our results indicate that starbursts are facilitated primarily by enhanced SFEs, the physical mechanism that leads to this increase remains unclear.  Galaxy interactions and mergers are commonly identified culprits as starburst triggers (e.g. Lambas et al. 2003; Nikolic et al. 2004; Woods et al. 2006; Ellison et al. 2008b, 2013; Li et al. 2008; Darg et al. 2010;  Patton et al. 2011, 2013; Scott \& Kaviraj 2014).  However, only three out of the 12 galaxies in our sample show obvious signs of a recent interaction.  Bars may also trigger central starbursts (Ellison et al. 2011; Carles et al. 2016; Martel et al. 2018), but the galaxies in our sample do not seem to be dominated by strong bars (Fig \ref{maps_fig}).  In practice, there are likely to be multiple mechanisms that provide the physical conditions necessary for triggering starbursts.  In addition to major mergers and bars, this could include accretion of small satellites (i.e. minor mergers, e.g. Woods \& Geller 2007; Scudder et al. 2012; Kaviraj 2014), interaction with an external medium (e.g. Vulcani et al. 2019b) or smooth accretion from the intergalactic medium (Sanchez-Almeida et al. 2014).

Ellison et al. (2008a) were the first to show that galaxies above the global star-forming main sequence have metallicities that are below the mass-metallicity relation, and suggested that this could indicate the role of gas accretion/metallicity dilution in the triggering of starbursts (see also Peeples et al. 2009)\footnote{Although it has become common to refer to this relationship as `fundamental', there are considerable subtleties involved.  For example, whilst mergers do qualitatively have the lower metallicities expected from their elevated SFRs (e.g. Kewley et al. 2006; Ellison et al. 2008b; Michel Dansac et al., 2008; Scudder et al. 2012), mergers are actually outliers in metallicity, given their SFRs (Gronnow et al. 2015; Bustamante et al. 2018, 2019).  The same is true for barred galaxies (Ellison et al. 2011).}.  The anti-correlation between star formation rates and metallicities that has been seen on global scales has recently been demonstrated on local (kpc) scales as well, with low metallicity regions exhibiting elevated SFRs (Ellison et al. 2018; Sanchez-Almeida et al. 2018; Hwang et al. 2019; Sanchez Menguiano et al. 2019).  These works have suggested that metal-poor gas inflows may be responsible for both diluting O/H and enhancing \sigsfr.  In such a model, gas flows onto, or within, galaxies dilute the gas phase metallicity in the interstellar medium prior to triggering a starburst.

In order to investigate whether the starbursts in our sample exhibit anomalously low metallicities, we compute an additional offset metric, \doh, which we have introduced previously in Ellison et al. (2018) and Thorp et al. (2019) and is analogous to \dsigsfr.  In brief, \doh\ is the offset of a given spaxel from the resolved mass-metallicity relation.  For every star-forming spaxel in our sample, we compute the difference of its log O/H from the median of all other star-forming spaxels matched in \sigstar, $R/R_e$ and total stellar mass.

In Fig. \ref{doh_profiles_fig} we show the \doh\ profiles of the 12 starburst galaxies in our sample.  Panels are plotted in the same order as Fig. \ref{all_profiles_fig}, although the y-axis range is zoomed in on Fig.  \ref{doh_profiles_fig}.  The \doh\ profiles are relatively flat, with deviations typically $<$ 0.05 dex. There are two galaxies with significantly offset central metallicities: 8156-3701 and 8081-3704.  These are the same two galaxies that show the strongest \dsfr\ and \dsfe\ enhancements in Fig.  \ref{all_profiles_fig} and have the greatest globally enhanced SFRs.  Recall that one of these galaxies is a post-merger, whereas the other does not show signs of a recent interaction.  The supressed central metallicities in these two galaxies are consistent with the results presented in Ellison et al. (2018), where we showed that starburst galaxies have low central metallicities in both interacting and non-interacting galaxies.  Although the sample is still small, the results suggest that metal-poor gas inflows from either merger-driven, or secular events can trigger starbursts.  Whilst one intuitive effect of inflows is to increase the gas fraction in the central regions, our analysis indicates that it is not \fh2\ that primarily drives the starburst, i.e. the galaxy's star formation does not simply scale up the normal Kennicutt-Schmidt relation.  Although gas fraction enhancements may play a role (e.g. Ellison et al. 2019), it appears that the dominant effect behind the starburst is an elevated SFE.  

In Fig. \ref{doh_fig} we again move beyond a galaxy-by-galaxy analysis and consider the spaxel population as a whole, plotting \doh\ vs. \dsigsfr, with colour coding according to \dsfe. Once again, we remind the reader that, due to the starburst selection, this figure is dominated by spaxels with positive \dsigsfr.    Fig. \ref{doh_fig} shows that there is a weak anti-correlation between \dsigsfr\ and \doh, consistent with other recent results (Sanchez-Almeida et al. 2018; Hwang et al. 2019; Sanchez Menguiano et al. 2019).  Notably, spaxels with the highest \dsigsfr\ have \doh\ that are particularly offset below the normal metallicity scaling relation.  As can be seen from the colour coding, these are the same spaxels that have particularly high \dsfe\ in Fig. \ref{correlation_fig}.   Taken together, our results indicate that, in regions with the strongest starbursts, metal poor gas injection has altered the ISM properties such that the SFE is increased.  

\section{Summary}\label{summary_sec}

In Ellison et al. (2018) we showed that galaxies located above the global SFMS preferentially have their star formation rates boosted in their central regions.  In the work presented here, we explore whether these central starbursts are primarily driven by an enhanced SFE, or whether it is an elevated gas fraction that leads to excess star formation.  We study 12 galaxies with central starbursts selected from the \aq\ survey (Lin et al., in prep) for which we have obtained ALMA maps of CO surface density on the same spatial scale as optical IFU data from MaNGA.  We compute offsets from known scaling relations to investigate which properties vary from their expected values: \dsigsfr, \dsfe\ and \dfgas\ are a given spaxel's offset from the rSFMS, the Kennicutt-Schmidt relation and the molecular gas main sequence, respectively.  Based on an analysis that studies averaged central properties and radial gradients, as well as the complete ensemble of $\sim$ 5300 spaxels in our sample, we find that the central starburst is primarily driven by enhanced star formation efficiencies.  Elevated gas fractions may play an additional role, but this is uncertain due to the likely variable CO conversion factors in the star bursting spaxels.  Star formation efficiencies are particularly enhanced in galaxies with the greatest offset from the global SFMS.  These most significant starburst galaxies also show central supressions in their O/H, consistent with a scenario in which the starburst has been triggered by an influx of low metallicity fuel.  Since only a minority (25 percent) of our sample show signs of a recent interaction, we conclude that a variety of mechanisms can be responsible for these central starbursts, although the generic characteristic is of an enhanced star formation efficiency.

\section*{Acknowledgements}

SLE acknowledges support from an NSERC Discovery Grant and thanks Erik Rosolowsky for discussions and advice throughout this project and Brenda Matthews for her patience in teaching an optical astronomer about long wavelengths and interferometry.  SLE, MDT, LL and HAP also gratefully acknowledge grant MOST 107-2119-M-001-024 and 108-2628-M-001 -001 -MY3 for travel funding that facilitated both the ALMA data reduction and analysis of \aq\ survey.  AFLB gratefully acknowledges the ERC Advanced Grant 695671 `QUENCH', and support from the STFC.

This paper makes use of the following ALMA data: ADS/JAO.ALMA\#2015.1.01225.S, ADS/JAO.ALMA\#2017.1.01093.S, ADS/JAO.ALMA\#2018.1.00558.S, ADS/JAO.ALMA\#2018.1.00541.S.   ALMA is a partnership of ESO (representing its member states), NSF (USA) and NINS (Japan), together with NRC (Canada), MOST and ASIAA (Taiwan), and KASI (Republic of Korea), in cooperation with the Republic of Chile. The Joint ALMA Observatory is operated by ESO, AUI/NRAO and NAOJ.

The National Radio Astronomy Observatory is a facility of the National Science Foundation operated under cooperative agreement by Associated Universities, Inc.

The SDSS is managed by the Astrophysical Research Consortium for the Participating Institutions. The Participating Institutions are the American Museum of Natural History, Astrophysical Institute Potsdam, University of Basel, University of Cambridge, Case Western Reserve University, University of Chicago, Drexel University, Fermilab, the Institute for Advanced Study, the Japan Participation Group, Johns Hopkins University, the Joint Institute for Nuclear Astrophysics, the Kavli Institute for Particle Astrophysics and Cosmology, the Korean Scientist Group, the Chinese Academy of Sciences (LAMOST), Los Alamos National Laboratory, the Max-Planck-Institute for Astronomy (MPIA), the Max-Planck-Institute for Astrophysics (MPA), New Mexico State University, Ohio State University, University of Pittsburgh, University of Portsmouth, Princeton University, the United States Naval Observatory, and the University of Washington.


\begin{thebibliography}{}
\small
\itemindent -0.48cm


\bibitem[Barrera-Ballesteros et al. (2016)]{jbb16} 
        Barrera-Ballesteros, J. K., et al., 2016, MNRAS, 463, 2513

\bibitem[Belfiore et al. (2017)]{bel17a}  
        Belfiore, F., et al., 2017, MNRAS, 469, 151

\bibitem[Bemis \& Wilson (2019)]{bw19}
        Bemis, A., \& Wilson, C., 2019, AJ, 157, 131
        
\bibitem[Bigiel et al. (2008)]{big08} 
        Bigiel, F., Leroy, A., Walter, F., Brinks, E., de Blok, W. J. G., 
	Madore, B., Thornley, M. D.,  2008, AJ, 136, 2846

\bibitem[Bigiel et al. (2011)]{big11} 
         Bigiel, F., et al., 2011, ApJ, 730, L13

\bibitem[Bigiel et al. (2016)]{big16} 
        Bigiel, F., et al. 2016, ApJ, 822, L26

\bibitem[Bluck et al. (2014)]{asa14}
        Bluck, A. F. L., Mendel, J. T., Ellison, S. L., Moreno, J., 
	Simard, L., Patton, D. R., Starkenburg, E.,
	2014, MNRAS, 441, 599

\bibitem[Bluck et al. (2016)]{asa16}
         Bluck, A. F. L., et al., 2016, MNRAS, 462, 2559

\bibitem[Bluck et al. (2019)]{bl19}
        Bluck, A. F. L.,  Maiolino, R., Sanchez, S. F., Ellison, S. L.,
        Thorp, M. D.,  Piotrowska, J., Teimoorinia, H., Bundy, K. A., MNRAS, 2019, submitted

\bibitem[Bolatto et al. (2017)]{bol17}
        Bolatto, A. D., et al. 2017, ApJ, 846, 159

\bibitem[Bolatto et al. (2013)]{bol13}
        Bolatto, A. D., Wolfire, M., Leroy, A. K., 2013, ARA\&A, 51, 207

\bibitem[Bothwell et al. (2013)]{both13}
        Bothwell, M. S., Maiolino, R., Kennicutt, R., Cresci, G., 
	Mannucci, F., Marconi, A., Cicone, C.,  2013, MNRAS, 433, 1425

\bibitem[Bothwell et al. (2016)]{both16}
        Bothwell, M. S., Maiolino, R., Peng, Y., Cicone, C., Griffith, H.,
	Wagg, J., 2016, MNRAS, 455, 1156

\bibitem[Brinchmann et al. 2004]{bri04} 
        Brinchmann, J., Charlot, S., White, S. D. M., Tremonti, C., 
	Kauffmann, G., Heckman, T., Brinkmann, J.,2004, MNRAS, 351, 1151  

\bibitem[Brown et al. (2018)]{brown18}
        Brown, T., Cortese, L., Catinella, B., Kilborn, V.,
	2018, MNRAS, 473, 1868

\bibitem[Bryant \& Scoville (1999)]{bs99}
         Bryant, P. M., \& Scoville, N., 1999, AJ, 117, 2632

\bibitem[Bundy et al. (2015)]{bun15}
        Bundy, K., et al., 2015, ApJ, 798, 7

\bibitem[Bustamante et al. (2018)]{bus18}
         Bustamante, S., Sparre, M., Springel, V., Grand, R. J. J.,
	 2018, MNRAS, 479, 3381

\bibitem[Bustamante et al. (2019)]{bus19}
         Bustamante, S., Ellison, S. L., Patton, D. R., Sparre, M.,
	 2019, MNRAS, submitted

\bibitem[Cano-Diaz et al. (2012)]{cd12}
        Cano-D\'{i}az, M., Maiolino, R., Marconi, A., Netzer, H., 
	Shemmer, O., Cresci, G., 2012, A\&A, 537, L8

\bibitem[Cardelli, Clayton \& Mathis (1989)]{ccm89}
        Cardelli, J. A., Clayton, G. C., \& Mathis, J. S., 1989, 
	ApJ, 345, 245

\bibitem[Carles et al. (2016)]{car16}
        Carles, C., Martel, H., Ellison, S. L., Kawata, D.,
	 2016, MNRAS, 463, 1074

\bibitem[Catinella et al. (2010)]{cat10}
        Catinella, B., et al,
	2010, MNRAS, 403, 683

\bibitem[Catinella et al. (2018)]{cat18}
        Catinella, B., et al, 2018, MNRAS, 476, 875

\bibitem[Chown et al. (2019)]{cho19}
        Chown, R., et al., 2019, MNRAS, 484, 5192

\bibitem[Cormier et al. (2018)]{cor18}
        Cormier, D., et al., 2018, MNRAS, 475, 3909

\bibitem[Daddi et al. (2010)]{dad10} 
        Daddi, E., et al.  2010, ApJ, 714, L118

\bibitem[Darg et al. (2010a)]{darg10a} 
        Darg, D. W., et al., 2010, MNRAS, 401, 1552


\bibitem[Downes \& Solomon (1998)]{ds98}
         Downes, D., \& Solomon, P. M., 1998, ApJ, 507, 615

\bibitem[Dutta et al. (2018)]{dutta18}
        Dutta, R., Srianand, R., Gupta, N., 2018, MNRAS, 480, 947

\bibitem[Dutta et al. (2019)]{dutta19}
        Dutta, R., Srianand, R., Gupta, N., 2019, MNRAS in press

\bibitem[Elbaz et al. (2018)]{elb18}
        Elbaz, D., et al., 2018, A\&A, 616, 110

\bibitem[Ellison et al., (2018)]{HI_in_PM}
         Ellison, S. L., Catinella, B., \& Cortese, L., 2018,
	 MNRAS, 478, 3447

\bibitem[Ellison et al., (2015)]{HI}
        Ellison, S. L., Fertig, D., Rosenberg, J. L., Nair, P., 
	Simard, L., Torrey, P., Patton, D. R., 2015, MNRAS, 448, 221

\bibitem[Ellison et al. (2013)]{post} 
         Ellison, S. L., Mendel, J. T.,  Patton, D. R., Scudder, J. M.,
	2013, MNRAS, 453, 3627

\bibitem[Ellison et al. (2011)]{bars}
         Ellison, S. L., Nair, P., Patton, D. R., Scudder, J. M., Mendel, 
	 J. T., Simard, L., 2011, MNRAS, 416, 2182

\bibitem[Ellison et al. (2008)]{sle08a} 
         Ellison, S. L., Patton, D. R., Simard, L., McConnachie, A. W.,
	 2008a, ApJ, 672, L107

\bibitem[Ellison et al. (2008)]{sle08b} 
         Ellison, S. L., Patton, D. R., Simard, L., McConnachie, A. W.,
	 2008b, AJ, 135, 1877

\bibitem[Ellison et al., (2018)]{manga}
        Ellison, S. L., Sanchez, S. F., Ibarra-Medel, H., Antonio, B.,
	Mendel, J. T., Barrera-Ballesteros, J., 2018, MNRAS, 474, 2039

\bibitem[Ellison et al. 2019]{ell19}
        Ellison, S. L., et al., 2019, MNRAS, in press
        
\bibitem[Espada et al. (2018)]{es18}
        Espada, D., et al. 2018, ApJ, 866, 77


\bibitem[Gallagher et al. (2018)]{gal18}
        Gallagher, M. J., et al., 2018, ApJ, 858, 90

\bibitem[Gao \& Solomon (2004)]{gs04} 
         Gao, Y., \& Solomon, P. M., 2004, ApJ, 606, 271 

\bibitem[Garcia-Burillo et al. (2012)]{gb12}
         Garcia-Burillo, S., Usero, A., Alonso-Herrero, A., Gracia-Carpio, J.,
	 Pereira-Santaella, M., Colina, L., Planesas, P., Arribas, S.,
	 2012, A\&A, 539, 8

\bibitem[Genzel et al. (2010)]{gen10}
         Genzel, R., et al. 2010, MNRAS, 407, 2091

\bibitem[Genzel et al. (2015)]{gen15}
        Genzel, R., et al., 2015, ApJ, 800, 20

\bibitem[Gonzalez-Delgado et al (2014)]{gd14}
        Gonzalez-Delgado, R., et al., 2014, A\&A, 562, 47

\bibitem[Gonzalez-Delgado et al (2015)]{gd15}
        Gonzalez-Delgado, R., et al., 2015, A\&A, 581, 103

\bibitem[Gracia-Carpio et al. (2008)]{gc08}
        Gracia-Carpio, J., Garcia-Burillo, S., Planesas, P., Fuente, A.,
	Usero, A., 2008, A\&A, 479, 703

\bibitem[Gronnow et al. 2015]{gro15}
        Gronnow, A. E., 2015, MNRAS, 451, 4005

\bibitem[Hani et al. (2019)]{hani19} 
        Hani, M. H., et al., 2019, MNRAS, submitted

\bibitem[Hoopes et al. (2007)]{hoo07}
        Hoopes, C. G., et al. 2007, ApJS, 2007, ApJS, 173, 441

\bibitem[Hsieh et al. (2017)]{hs17}
        Hsieh, B. C., et al., 2017, ApJ, 851, L24

\bibitem[Hwang et al. (2019)]{hwa19}
         Hwang, H.-C., et al., 2019, ApJ, 872, 144

\bibitem[Jimenez-Donaire et al. (2019)]{jd19}
        Jimenez-Donaire, M. A., et al., 2019, ApJS, in press 
        
\bibitem[Kauffmann et al. (2003)]{kau03b} 
	Kauffmann, G., et al., 2003, MNRAS, 346, 1055

\bibitem[Kaviraj (2014)]{kav14}
        Kaviraj, S., 2014, MNRAS, 440, 2944

\bibitem[Kennicutt (1998)]{ken98a}
        Kennicutt, R.C. 1998a, ARA\&A, 36, 189

\bibitem[Kennicutt (1989)]{ken89}
         Kennicutt, R.C., 1989, ApJ, 344, 685

\bibitem[Kennicutt (1998)]{ken98b}
         Kennicutt, R.C., 1998b, ApJ, 498, 541

\bibitem[Kewley \& Ellison (2008)]{ke08}
        Kewley, L. J., \& Ellison, S. L., 2008, ApJ, 681, 1183

\bibitem[Kewley et al. (2006)]{kewl06}
         Kewley, L. J., Geller, M. J., Barton, E. J.,
	  2006, AJ, 131, 2004

\bibitem[Lada, Lombardi \& Alves (2010)]{lla10}
        Lada, C. J., Lombardi, M.,  \& Alves, J. F.,  2010, ApJ, 724, 687

\bibitem[Lambas et al. (2003)]{lam03}
        Lambas, D. G., Tissera, P. B., Alonso, M. S., Coldwell, G.,
	2003, MNRAS, 346, 1189

\bibitem[Lara-Lopez et al. (2010)]{ll10} 
        Lara-Lopez, M. A., et al., 2010, A\&A, 521, L53

\bibitem[Lee et al. (2017)]{lee17}
         Lee, N., et al., 2017, MNRAS, 471, 2124

\bibitem[Leroy et al. (2008)]{ler08} 
        Leroy, A. K., Walter, F., Brinks, E., Bigiel, F., de Blok, W. J. G.,
	Madore, B., Thornley, M. D., 2008, AJ, 136, 2782

\bibitem[Leroy et al. (2009)]{ler09} 
	Leroy, A. K., et al., 2009, AJ, 137, 4670

\bibitem[Leroy et al. (2013)]{ler13} 
        Leroy, A. K., et al., 2013, AJ, 146, 19

\bibitem[Li et al. (2008a)]{li08a} 
        Li, C., Kauffmann, G., Heckman, T. M., Jing, Y. P., White, S. D. M.,
	2008, MNRAS, 385, 1903

\bibitem[Lin et al. (2014)]{lin14} 
        Lin, L., et al., 2014, ApJ, 782, 33
          
\bibitem[Lin et al. (2019)]{lin19} 
        Lin, L., et al., 2019, ApJ, 884, L33

\bibitem[Magdis et al. (2012)]{mag12}
         Magdis, G. E., et al.,  2012, ApJ, 760, 6

\bibitem[Mannucci et al. (2010)]{man10}
       Mannucci, F., Cresci, G., Maiolino, R., Marconi, A., Gnerucci, A.,
       2010, MNRAS, 408, 2115

\bibitem[Marino et al. (2013)]{mar13}
        Marino, R. A., et al.,  2013, A\&A, 559A, 114

\bibitem[Martel et al., (2018)]{hm18}
        Martel, H.,  Carles, C., Robichaud, F., Ellison, S. L., Williamson,
	D. J.,  2018, MNRAS, 477, 5367

\bibitem[Michel-Dansac et al. (2008)]{md08}
        Michel-Dansac, L., Lambas, D. G., Alonso, M. S., Tissera, P.,
	2008, MNRAS, 386, 82


\bibitem[Narayanan et al. (2011)]{nar11}
         Narayanan, D., Krumholz, M., Ostriker, E. C., Hernquist, L.,
	  2011, MNRAS, 418, 664

\bibitem[Nikolic, Cullen \& Alexander (2004)]{nca04}
        Nikolic, B., Cullen, H., Alexander, P., 2004, MNRAS, 355, 874

\bibitem[Noeske et al. (2007)]{noe07}
        Noeske, K. G., et al.,  2007, ApJ, 660, L43

\bibitem[Ostriker, McKee, Leroy (2010)]{oml10}
        Ostriker, E. C., McKee, C. F., Leroy, A. K., 2010, ApJ, 721, 975

\bibitem[Pan et al. (2018)]{pan18}
        Pan, H.-A., et al., 2018, ApJ, 868, 132
 
\bibitem[Pan et al. (2019)]{pan19} 
        Pan, H.-A., et al., 2019, ApJ, 881, 119

\bibitem[Papadopoulos et al. 2012]{pap12}
         Papadopoulos, P. P., van der Werf, P., Xilouris, E., Isaak, K. G.,
	 Gao, Y., 2012, ApJ, 751, 10

\bibitem[Patton et al. (2011)]{dave11}
         Patton, D. R., Ellison, S. L.,  Simard, L., McConnachie, A. W.,
	 Mendel, J. T.,
	 2011, MNRAS, 412, 591 

\bibitem[Patton et al. (2013)]{dave13}
        Patton, D. R., Torrey, P., Ellison, S. L., Mendel, J. T., 
	Scudder, J. M., 2013, MNRAS, 433, L59

\bibitem[Peeples et al. (2009)]{peep09} 
        Peeples, M. S., Pogge, R. W., Stanek, K. Z., 2009, ApJ, 695, 259

\bibitem[Peng et al. (2010)]{peng10}
        Peng, Y., et al., 2010, ApJ, 721, 193

\bibitem[Rahman et al. (2012)]{rah12}
         Rahman, N., et al., 2012, ApJ, 745, 183



\bibitem[Saintonge et al. (2011)]{sain11}
        Saintonge, A., et al., 2011, MNRAS, 415, 61

\bibitem[Saintonge et al. (2012)]{sain12}
         Saintonge, A., et al., 2012, ApJ, 758, 73

\bibitem[Saintonge (2016)]{sain16}
        Saintonge, A., et al., 2016, MNRAS, 462, 1749

\bibitem[Saintonge (2017)]{sain17}
        Saintonge, A., et al., 2017, ApJS, 233, 22

\bibitem[Salim et al. (2007)]{sal07}
        Salim, S., et al, 2007, ApJS, 173, 267

\bibitem[Salim et al. (2014)]{sal14}
        Salim, S., Lee, J. C., Ly, C., Brinchmann, J., Dave, R.,
	Dickinson, M., Salzer, J. J., Charlot, S., 2014, ApJ, 797, 126

\bibitem[S\'{a}nchez et al. (2012)]{s12}
         S\'{a}nchez, S. F., et al., 2012, A\&A, 538, 8

\bibitem[S\'{a}nchez et al. (2013)]{s13}
         S\'{a}nchez, S. F., et al., 2013, A\&A, 554, 58

\bibitem[S\'{a}nchez et al. (2016a)]{pipe16a}
        S\'{a}nchez, S. F., et al., 2016a, RMxAA, 52, 21

\bibitem[S\'{a}nchez et al. (2016b)]{pipe16b}
        S\'{a}nchez, S. F., et al., 2016b, RMxAA, 52, 171

\bibitem[S\'{a}nchez et al. (20118)]{san18dr15}
        S\'{a}nchez, S. F., et al., 2018, RMxAA, 54, 217

\bibitem[Sanchez-Almeida et al. (2014)]{sa14}
         Sanchez Almeida, J., Elmegreen, B. G., Munoz-Tunon, C., Elmegreen, 
         D. M., 2014, A\&AR, 22, 71

\bibitem[Sanchez-Almeida et al. (2018)]{sa18}
        Sanchez Almeida, J., Caon, N., Munoz-Tunon, C., Filho, M., Cervino, M.,
	 2018, MNRAS, 476, 4765 

\bibitem[Sanchez-Menguiano et al. (2018)]{sm18}
        Sanchez-Menguiano, L., et al., 2018, A\&A, 609, 119 

\bibitem[Sanchez-Menguiano et al. (2019)]{sm19}
         Sanchez-Menguiano, L., Sanchez Almeida, J., Munoz-Tunon, C.,
	 Sanchez, S. F., Filho, M., Hwang, H.-.C, Drory, N.,
	 2019, ApJ, in press


\bibitem[Sandstrom et al. (2013)]{sand11}
        Sandstrom, K. M., et al., 2013, ApJ, 777, 5

\bibitem[Sargent et al. (2014)]{sar14}
         Sargent, M. T., et al. 2014, ApJ, 793, 19

\bibitem[Schmidt (1959)]{sch59}
        Schmidt, M., 1959, ApJ, 129, 243

\bibitem[Schruba et al. (2011)]{sch11}
        Schruba, A., et al., 2011, AJ, 142, 37

\bibitem[Scott \& Kaviraj (2014)]{sk14}
         Scott, C., \& Kaviraj, S., 2014, MNRAS, 437, 2137

\bibitem[Scoville et al. (2016)]{sco16}
       Scoville, N., et al.,  2016, ApJ, 820, 83

\bibitem[Scoville et al. (2017)]{sco17}
        Scoville, N., et al.,  2017, ApJ, 837, 150

\bibitem[Scudder et al. (2012b)]{[pairs}
         Scudder, J. M., Ellison, S. L., Torrey, P., Patton, D. R.,
	 Mendel, J. T., 2012, MNRAS, 426, 549

\bibitem[Silverman et al. (2015)]{sil15}
	Silverman, J.D., et al., 2015, ApJ, 812, L23

\bibitem[Silverman et al. (2018)]{sil18}
         Silverman, J.D., et al., 2018, ApJ, 867, 92

\bibitem[Solomon \& Sage (1988)]{ss88}
        Solomon, P. M., \& Sage, L. J.,  1988, ApJ, 334, 613


\bibitem[Tacconi et al. (2018)]{tac18}
        Tacconi, L. J., et al., 2018, ApJ, 853, 179

\bibitem[Teimoorinia, Bluck \& Ellison (2016)]{tbe16}
        Teimoorinia, H., Bluck, A. F. L., \& Ellison, S. L.,
	 2016, MNRAS, 457, 2086

\bibitem[Thorp et al. (2019)]{tho19}
        Thorp, M. D., Ellison, S. L., Simard, L., Sanchez, S. F.,
	Antonio, B.,  2019, MNRAS, 482, L55

\bibitem[Tomicic et al. (2019)]{tom19}
        Tomicic, N., et al., 2019, ApJ, 873, 3
	 
\bibitem[Tremonti et al. 2004]{tre04} 
        Tremonti, C., et al., 2004, ApJ, 693, 898

\bibitem[Ueda et al. (2014)]{ueda14}
         Ueda, J., et al. 2014, ApJS, 214, 1

\bibitem[Usero et al., (2015)]{us15}
        Usero, A., et al., 2015, AJ, 150, 115

\bibitem[Utomo et al. (2017)]{uto17} 
        Utomo, D., et al., 2017, ApJ, 849, 26


\bibitem[Violino et al. 2018]{vio}
        Violino, G., Ellison, S. L., Sargent, M., Coppin, K. E. K., 
	Scudder, J. M., Mendel, J. T.,  Saintonge, A.,
	MNRAS, 2018, 476, 2591

\bibitem[Vulcani et al. (2019)]{vul19} 
        Vulcani, B., et al., 2019a, MNRAS, in press

\bibitem[Vulcani et al. (2019)]{igm}
        Vulcani, B., et al., 2019b, MNRAS, 487, 2278

\bibitem[Wang et al. (2019)]{wang19}
        Wang, E., Lilly, S. J., Pezzulli, G., Matthee, J.,
	 2019, ApJ, 877, 132

\bibitem[Woo et al. (2013)]{woo13}
        Woo, J., et al., 2013, MNRAS, 428, 3306

\bibitem[Woo et al. (2015)]{woo15}
        Woo, J., Dekel, A., Faber, S. M., Koo, D. C., 2015, MNRAS, 448, 237

\bibitem[Woods \& Geller (2007)]{wg07}
        Woods, D. F., Geller, M. J., 
	2007, AJ, 134, 527

\bibitem[Woods et al. (2006)]{wgb06}
        Woods, D. F., Geller, M. J., Barton, E. J.,
	2006, AJ, 132, 197

\bibitem[Wong et al. (2013)]{wong13}
        Wong, T., et al., 2013, ApJ, 777, L4

\bibitem[Wu et al. (2005)]{wu05}
         Wu, J., Evans, N. J., Gao, Y., Solomon, P. M., Shirley, Y. L.,
	 Vanden Bout, P. A., 2005, ApJ, 635, 173

\bibitem[Wuyts et al. (2011)]{wuy11b}
         Wuyts, S., et al., 2011, ApJ, 742, 96



         
\end{thebibliography}
\end{document}